%% file: ms.tex
\documentclass{article}
\usepackage{arxiv}

\usepackage[utf8]{inputenc} 
\usepackage[T1]{fontenc}    
\usepackage{hyperref}       
\usepackage{url}            
\usepackage{booktabs}       
\usepackage{amsfonts}       
\usepackage{nicefrac}       
\usepackage{microtype}      
\usepackage{lipsum}		
\usepackage{graphicx}
\usepackage[authoryear]{natbib}
\usepackage{amsmath}
\usepackage{mathrsfs}
\usepackage{multirow}
\usepackage{pdflscape}

\newlength{\colwidth}
\title{Earthquake magnitude and location estimation from real time seismic waveforms with a transformer network}

\author{Jannes Münchmeyer$^{1, 2, *}$, Dino Bindi$^{1}$, Ulf Leser$^{2}$, Frederik Tilmann$^{1, 3}$\\
$^1$ Deutsches GeoForschungsZentrum GFZ, Potsdam, Germany\\
$^2$ Institut für Informatik, Humboldt-Universität zu Berlin, Berlin, Germany\\
$^3$ Insitut für geologische Wissenschaften, Freie Universität Berlin, Berlin, Germany\\
$^*$ To whom correspondence should be addressed: \url{munchmej@gfz-potsdam.de}
}

\renewcommand{\headeright}{}
\renewcommand{\shorttitle}{Earthquake magnitude and location estimation with TEAM-LM}

\begin{document}
\maketitle

\begin{abstract}
Precise real time estimates of earthquake magnitude and location are essential for early warning and rapid response.
While recently multiple deep learning approaches for fast assessment of earthquakes have been proposed, they usually rely on either seismic records from a single station or from a fixed set of seismic stations.
Here we introduce a new model for real-time magnitude and location estimation using the attention based transformer networks.
Our approach incorporates waveforms from a dynamically varying set of stations and outperforms deep learning baselines in both magnitude and location estimation performance.
Furthermore, it outperforms a classical magnitude estimation algorithm considerably and shows promising performance in comparison to a classical localization algorithm.
Our model is applicable to real-time prediction and provides realistic uncertainty estimates based on probabilistic inference.
In this work, we furthermore conduct a comprehensive study of the requirements on training data, the training procedures and the typical failure modes.
Using three diverse and large scale data sets, we conduct targeted experiments and a qualitative error analysis.
Our analysis gives several key insights.
First, we can precisely pinpoint the effect of large training data; for example, a four times larger training set reduces average errors for both magnitude and location prediction by more than half, and reduces the required time for real time assessment by a factor of four.
Second, the basic model systematically underestimates large magnitude events.
This issue can be mitigated, and in some cases completely resolved, by incorporating events from other regions into the training through transfer learning.
Third, location estimation is highly precise in areas with sufficient training data, but is strongly degraded for events outside the training distribution, sometimes producing massive outliers.
Our analysis suggests that these characteristics are not only present for our model, but for most deep learning models for fast assessment published so far.
They result from the black box modeling and their mitigation will likely require imposing physics derived constraints on the neural network.
These characteristics need to be taken into consideration for practical applications.
\end{abstract}

\section{Introduction}

Recently, multiple studies investigated deep learning on raw seismic waveforms for the fast assessment of earthquake parameters, such as magnitude (e.g. \cite{lomaxInvestigationRapidEarthquake2019}, \cite{mousaviMachineLearningApproachEarthquake2020}, \cite{vandenendeAutomatedSeismicSource2020a}), location (e.g. \cite{kriegerowskiDeepConvolutionalNeural2019}, \cite{mousaviBayesianDeepLearningEstimationEarthquake2019}, \cite{vandenendeAutomatedSeismicSource2020a}) and peak ground acceleration (e.g. \cite{jozinovicRapidPredictionEarthquake2020}).
Deep learning is well suited for these tasks, as it does not rely on manually selected features, but can learn to extract relevant information from the raw input data.
This property allows the models to use the full information contained in the waveforms of an event.
However, the models published so far use fixed time windows and can not be applied to data of varying length without retraining.
Similarly, except the model by \cite{vandenendeAutomatedSeismicSource2020a}, all models process either waveforms from only a single seismic station or rely on a fixed set of seismic stations defined at training time.
The model by \cite{vandenendeAutomatedSeismicSource2020a} enables the use of a variable station set, but combines measurements from multiple stations using a simple pooling mechanism.
While it has not been studied so far in a seismological context, it has been shown in the general domain that set pooling architectures are in practice limited in the complexity of functions they can model \citep{lee2019set}.

Here we introduce a new model for magnitude and location estimation based on the architecture recently introduced for the transformer earthquake alerting model (TEAM) \citep{munchmeyerTransformerEarthquakeAlerting2020}, a deep learning based earthquake early warning model.
While TEAM estimated PGA at target locations, our model estimates magnitude and hypocentral location of the event.
We call our adaptation TEAM-LM, TEAM for location and magnitude estimation.
We use TEAM as a basis due to its flexible multi-station approach and its ability to process incoming data effectively in real-time, issuing updated estimates as additional data become available.
Similar to TEAM, TEAM-LM uses mixture density networks to provided probability distributions rather than merely point estimates as predictions.
For magnitude estimation, our model outperforms two state of the art baselines, one using deep learning \citep{vandenendeAutomatedSeismicSource2020a} and one classical approach \citep{kuyukGlobalApproachProvide2013}.
For location estimation, our model outperforms a deep learning baseline \citep{vandenendeAutomatedSeismicSource2020a} and shows promising performance in comparison to a classical localization algorithm.

We note a further deficiency of previous studies for deep learning in seismology.
Many of these pioneering studies focused their analysis on the average performance of the proposed models.
Therefore, little is known about the conditions under which these models fail, the impact of training data characteristics, the possibility of sharing knowledge across world regions, and of specific training strategies.
All of these are of particular interest when considering practical application of the models.

To address these issues and provide guidance for practitioners, we perform a comprehensive evaluation of TEAM-LM on three large and diverse data sets: a regional broadband data set from Northern Chile, a strong motion data set from Japan, and another strong motion data set from Italy.
These data sets differ in their seismotectonic environment (North Chile and Japan: subduction zones; Italy: dominated by both convergent and divergent continental deformation), their spatial extent (North Chile: regional scale; Italy and Japan: national catalogs), and the instrument type (North Chile: broadband, Italy and Japan: strong motion).
All three data sets contain hundreds of thousands of waveforms.
North Chile is characterized by a relatively sparse station distribution, but a large number of events and a low magnitude of completeness. 
There are far more stations in the Italy and Japan data sets, but a smaller number of earthquakes.
This selection of diverse data sets allows for a comprehensive analysis, giving insights for different use cases.
Our targeted experiments show that the characteristics are rooted in the principle structure used by TEAM-LM, i.e., the black box approach of learning a very flexible model from data, without imposing any physical constraints.
As this black box approach is common to all current fast assessment models using deep learning, they can be transferred to these models.
This finding is further backed by comparison to the results from previous studies.

\section{Data and Methods}

\subsection{Datasets}

\begin{figure}
 \includegraphics[width=\textwidth]{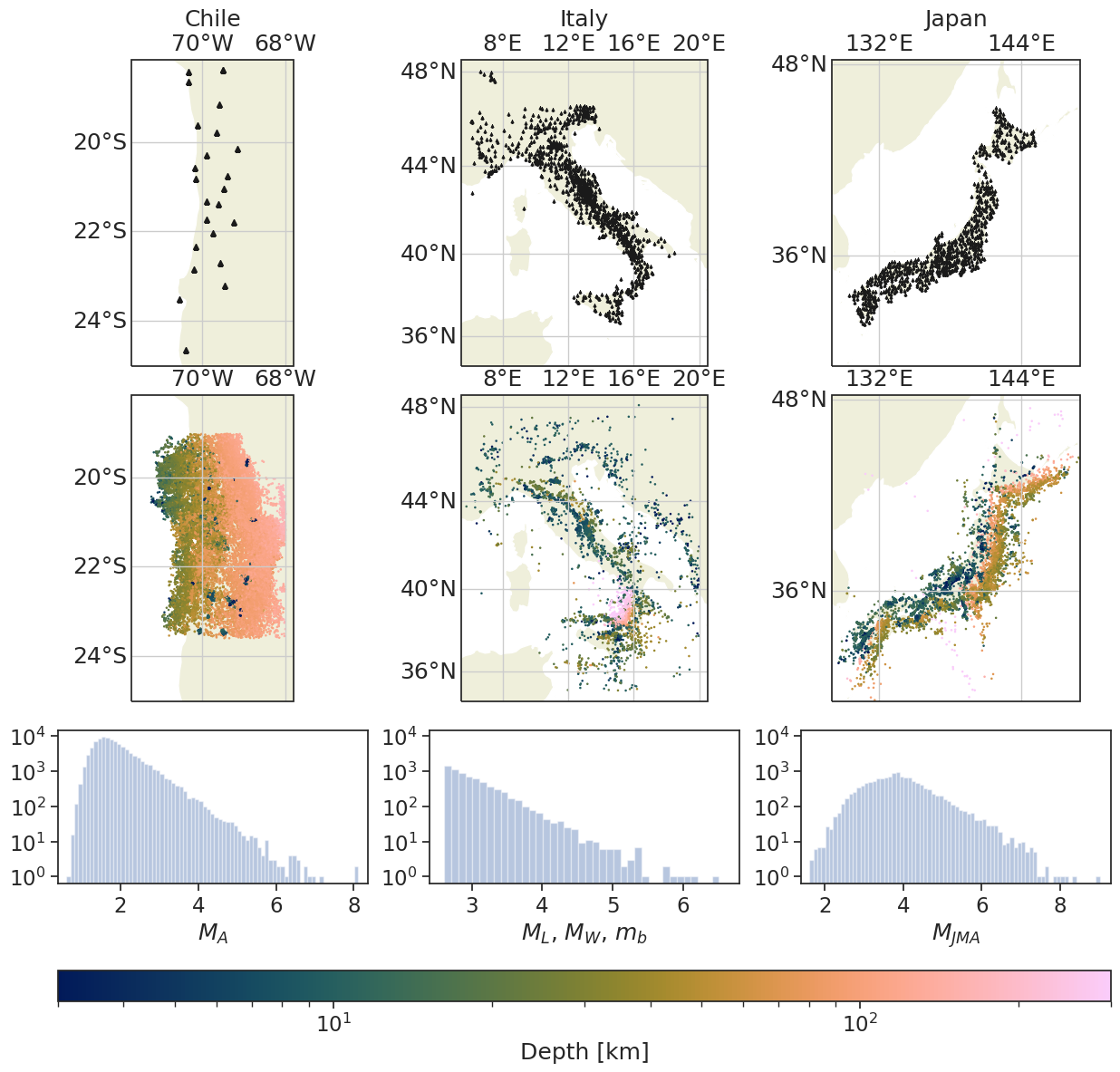}
 \caption{Overview of the data sets. The top row shows the spatial station distribution, the second tow the spatial event distribution. The event depth is encoded using color. Higher resolution versions of the maps can be found in the supplementary material (Figures \ref{SM-fig:highres_chile}, \ref{SM-fig:highres_italy}, \ref{SM-fig:highres_japan}). The bottom row shows the distributions of the event magnitudes. The magnitude scales are the peak displacement based $M_A$, local magnitude $M_L$, moment magnitude $M_W$, body wave magnitude $m_b$ and $M_\mathrm{JMA}$, a magnitude primarily using peak displacement.}
 \label{fig:data sets}
\end{figure}

\begin{table}
 \centering
 \caption{Overview of the data sets. The lower boundary of the magnitude category is the 5th percentile of the magnitude; this limit is chosen as each data set contains a small number of unrepresentative very small events.
 The upper boundary is the maximum magnitude. Magnitudes are given with two digit precision for Chile, as the precision of the underlying catalog is higher than for Italy and Japan.  The Italy data set uses different magnitudes for different events, which are $M_L$ ($>$90~\% of the events), $M_W$ ($<$10~\%) and $m_b$ ($<$1~\%).
 For depth and distance minimum, median and maximum are stated.
 Distance refers to the epicentral distance between stations and events. Note that the count of traces refers to the number of waveform-triplets (for three components, or group of six waveforms for the Japanese stations).
The sensor types are broadband (BB) and strong motion (SM).}
\input{datasets.tex}
\label{tab:data sets}
\end{table}

For this study we use three data sets (Table \ref{tab:data sets}, Figure \ref{fig:data sets}): one from Northern Chile, one from Italy and one from Japan.
The Chile data set is based on the catalog by \cite{sipplSeismicityStructureNorthern2018b} with the magnitude values from \cite{munchmeyerLowUncertaintyMultifeature2020}.
While there were minor changes in the seismic network configuration during the time covered by the catalog, the station set used in the construction of this catalog had been selected to provide a high degree of stability of the locations accuracy throughout the observational period \citep{sipplSeismicityStructureNorthern2018b}.
Similarly, the magnitude scale has been carefully calibrated to achieve a high degree of consistency in spite of significant variations of attenuation \citep{munchmeyerLowUncertaintyMultifeature2020}.
This data set therefore contains the highest quality labels among the data sets in this study.
For the Chile data set, we use broadband seismogramms from the fixed set of 24 stations used for the creation of the original catalog and magnitude scale.
Although the Chile data set has the smallest number of stations of the three data sets, it comprises three to four times as many waveforms as the other two due to the large number of events.

The data sets for Italy and Japan are more focused on early warning, containing fewer events and only strong motion waveforms.
They are based on catalogs from the INGV \citep{ingv2007} and the NIED KiKNet \citep{nied2019kiknet}, respectively.
The data sets each encompass a larger area than the Chile data set and include waveforms from significantly more stations.
In contrast to the Chile data sets, the station coverage differs strongly between different events, as only stations recording the event are considered.
In particular, KiKNet stations do not record continuous waveforms, but operate in trigger mode, only saving waveforms if an event triggered at the station.
For Japan each station comprises two sensors, one at the surface and one borehole sensor.
Therefore for Japan we have 6 component recordings (3 surface, 3 borehole) available instead of the 3 component recordings for Italy and Chile.
A full list of seismic networks used in this study can be found in the appendix (Table \ref{tab:seismic_networks}).

For each data set we use the magnitude scale provided in the catalog.
For the Chile catalog, this is $M_A$, a peak displacement based scale, but without the Wood-Anderson response and therefore saturation-free for large events \citep{munchmeyerLowUncertaintyMultifeature2020,deichmannWhyDoesML2018}.
For Japan $M_\mathrm{JMA}$ is used.
$M_\mathrm{JMA}$ combines different magnitude scales, but similarly to $M_A$ primarily uses horizontal peak displacement \citep{doi2014seismic}.
For Italy the catalog provides different magnitude types approximately dependent on the size of the event: $M_L$ ($>$90~\% of the events), $M_W$ ($<$10~\%) and $m_b$ ($<$1~\%).
We note that while the primary magnitude scales for all data sets are peak-displacement based, the precision of the magnitudes vary, with the highest precision for Chile.
This might lead to slightly worse magnitude estimation performance for Italy and Japan.

For all data sets the data were not subselected based on the type of seismicity but only based on the location (for Chile and Italy) or depending if they triggered (Japan).
This guarantees that, even though we made use of a catalog to assemble our training data, the resulting data sets are suitable for training and assessing methods geared at real-time applications without any prior knowledge about the earthquakes.
We focus on earthquake characterization and do not discuss event detection or separation from noise; we refer the interested reader to \citet{perolConvolutionalNeuralNetwork2018,mousaviCREDDeepResidual2018}.


We split each data set into training, development and test set.
For Chile and Japan we apply a simple chronological split with approximate ratios of 60:10:30 between training, development and test set, with the most recent events in the test set. 
As the last 30\% of the Italy data set consist of less interesting events for early warning, we instead use all events from 2016 as test set and the remaining events as training and development sets. 
We reserve all of 2016 for testing, as it contains a long seismic sequence in central Italy with two mainshocks in August ($M_W=6.5$) and October ($M_W=6.0$).
Notably, the largest event in the test set is significantly larger than the largest event in the training set ($M_w = 6.1$ L'Aquila event in 2007), representing a challenging test case.
For Italy, we assign the remaining events to training and development set randomly with a 6:1 ratio.

\subsection{The transformer earthquake alerting model for magnitude and location}

\begin{figure}
 \includegraphics[width=\textwidth]{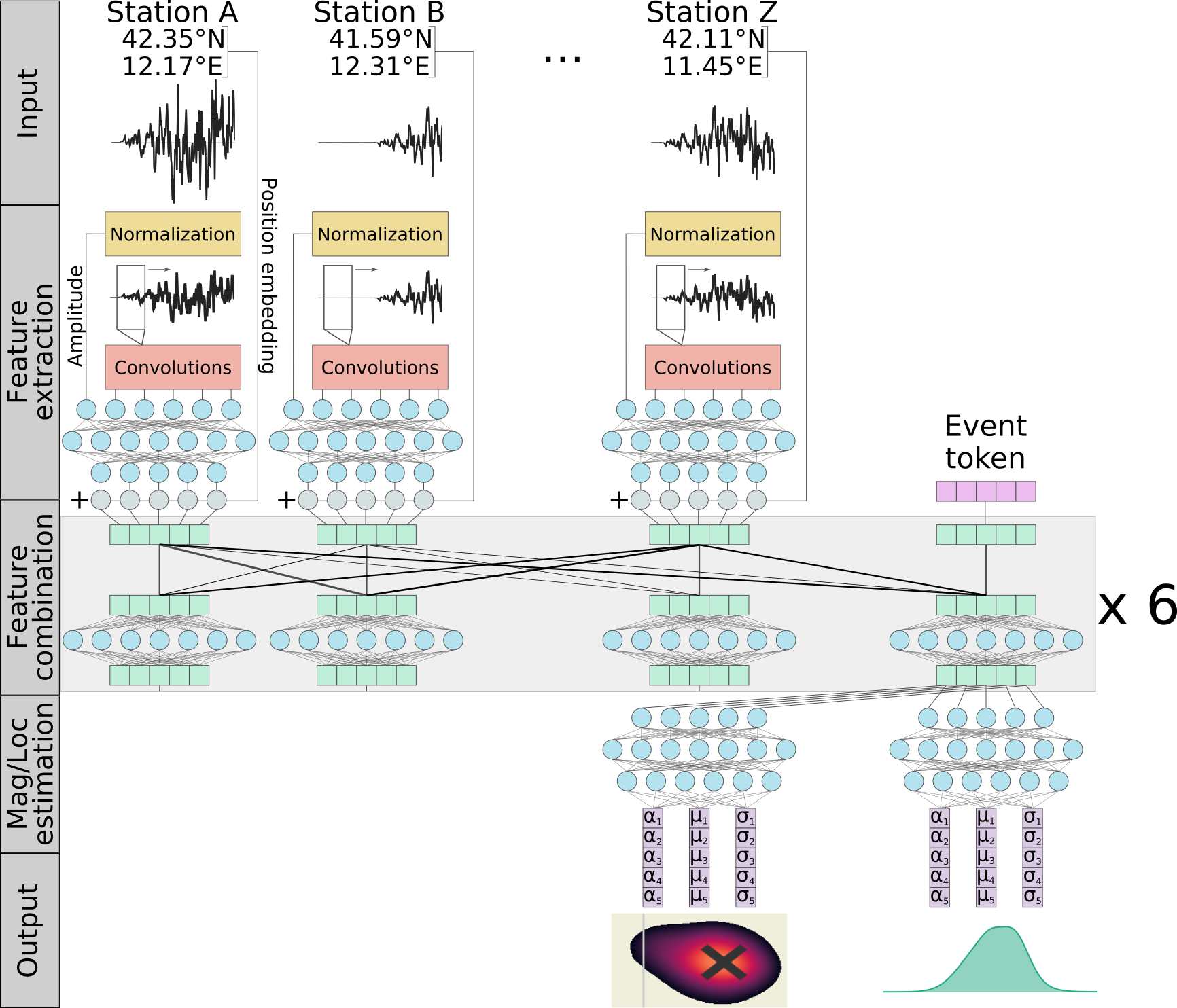}
  \caption{Overview of the adapted transformer earthquake alerting model, showing the input, the feature extraction, the feature combination, the magnitude/location estimation and the output. For simplicity, not all layers are shown, but only their order and combination is visualized schematically. For the exact number of layers and the size of each layer please refer to tables~\ref{SM-tab:convolutions} to \ref{SM-tab:mixture}. Please note that the number of input stations is variable, due to the self-attention mechanism in the feature combination.}
 \label{fig:model}
\end{figure}

We build a model for real time earthquake magnitude and location estimation based on the core ideas of the transformer earthquake alerting model (TEAM), as published in \citet{munchmeyerTransformerEarthquakeAlerting2020}.
TEAM is an end-to-end peak ground acceleration (PGA) model calculating probabilistic PGA estimates based on incoming waveforms from a flexible set of stations.
It employs the transformer network method \citep{vaswani2017attention}, an attention based neural network which was developed in the context of natural language processing (NLP), at the core of its algorithm.
Here, we adapt TEAM to calculate real time probabilistic estimates of event magnitude and hypocentral location.
As our model closely follows the architecture and key ideas of TEAM, we use the name TEAM-LM to refer to the location and magnitude estimation model.

Similar to TEAM, TEAM-LM consists of three major components (Figure \ref{fig:model}): a feature extraction, which generates features from raw waveforms at single stations, a feature combination, which aggregates features across multiple stations, and an output estimation.
Here, we briefly discuss the core ideas of the TEAM architecture and training and put a further focus on the necessary changes for magnitude and location estimation.
For a more detailed account of TEAM and TEAM-LM we refer to \cite{munchmeyerTransformerEarthquakeAlerting2020}, Tables \ref{SM-tab:convolutions} to \ref{SM-tab:mixture} and the published implementation.

The input to TEAM consists of three component seismogramms from multiple stations and their locations.
TEAM aligns all seismogramms to start and end at the same times $t_0$ and $t_1$.
We choose $t_0$ to be 5 seconds before the first P arrival at any station.
This allows the model to understand the noise conditions at all stations.
We limit $t_1$ to be at latest $t_0 + 30$~s.
In a real-time scenario $t_1$ is the current time, i.e., the available amount of waveforms, and we use the same approach to imitate real-time waveforms in training and evaluation.
The waveforms are padded with zeros to a length of 30~s to achieve constant length input to the feature extraction.

TEAM uses a CNN architecture for feature extraction, which is applied separately at each station.
The architecture consists of several convolution and pooling layers, followed by a multi-layer perceptron (Table \ref{SM-tab:convolutions}).
To avoid scaling issues, each input waveform is normalized through division by its peak amplitude.
As the amplitude is expected to be a key predictor for the event magnitude, we provide the logarithm of the peak amplitude as a further input to the multi-layer perceptron inside the feature extraction network.
We ensure that this transformation does not introduce a knowledge leak by calculating the peak amplitude only based on the waveforms until $t_1$.
The full feature extraction returns one vector for each station, representing the measurements at the station.

The feature vectors from multiple stations are combined using a transformer network \citep{vaswani2017attention}.
Transformers are attention based neural networks, originally introduced for natural language processing.
A transformer takes a set of $n$ vectors as input, and outputs again $n$ vectors which now incorporate the context of each other.
The attention mechanism allows the transformer to put special emphasis on inputs that it considers particularly relevant and thereby model complex inter-station dependencies.
Importantly, the parameters of the transformer are independent of the number of input vectors $n$, allowing to train and apply a transformer on variable station sets.
To give the transformer a notion of the position of the stations, TEAM encodes the latitude, longitude and elevation of the stations using a sinusoidal embedding and adds this embedding to the feature vectors.

TEAM adds the position embeddings of the PGA targets as additional inputs to the transformer.
In TEAM-LM, we aim to extract information about the event itself, where we do not know the position in advance.
To achieve this, we add an event token, which is a vector with the same dimensionality as the positional embedding of a station location, and which can be thought of as a query vector.
This approach is inspired by the so-called sentence tokens in NLP that are used to extract holistic information on a sentence \citep{devlin2018bert}.
The elements of this event query vector are learned during the training procedure.

From the transformer output, we only use the output corresponding to the event token, which we term event embedding and which is passed through another multi-layer perceptron predicting the parameters and weights of a mixture of Gaussians \citep{bishop1994mixture}.
We use $N=5$ Gaussians for magnitude and $N=15$ Gaussians for location estimation.
For computational and stability reasons, we constrain the covariance matrix of the individual Gaussians for location estimation to a diagonal matrix to reduce the output dimensionality.
Even though uncertainties in latitude, longitude and depth are known to generally be correlated, this correlation can be modeled with diagonal covariance matrices by using the mixture.


The model is trained end-to-end using a log-likelihood loss with the Adam optimizer \citep{kingma2014adam}.
We train separate models for magnitude and for location.
As we observed difficulties in the onset of the optimization when starting from a fully random initialization, we pretrain the feature extraction network.
To this end we add a mixture density network directly after the feature extraction and train the resulting network to predict magnitudes from single station waveforms.
We then discard the mixture density network and use the weights of the feature extraction as initialization for the end-to-end training.
We use this pretraining method for both magnitude and localization networks. 

Similarly to the training procedure for TEAM we make extensive use of data augmentation during training.
First, we randomly select a subset of up to 25 stations from the available station set.
We limit the maximum number to 25 for computational reasons.
Second, we apply temporal blinding, by zeroing waveforms after a random time $t_1$.
This type of augmentation allows TEAM-LM to be applied to real time data.
We note that this type of temporal blinding to enable real time predictions would most likely work for the previously published CNN approaches as well.
To avoid knowledge leaks for Italy and Japan, we only use stations as inputs that triggered before time $t_1$ for these data sets.
This is not necessary for Chile, as there the maximum number of stations per event is below 25 and waveforms for all events are available for all stations active at that time, irrespective of whether the station actually recorded the event. 
Third, we oversample large magnitude events, as they are strongly underrepresented in the training data set.
We discuss the effect of this augmentation in further detail in the Results section.
In contrast to the station selection during training, in evaluation we always use the 25 stations picking first.
Again, we only use stations and their waveforms as input once they triggered, thereby ensuring that the station selection does not introduce a knowledge leak.

\subsection{Baseline methods}

Recently, \cite{vandenendeAutomatedSeismicSource2020a} suggested a deep learning method capable of incorporating waveforms from a flexible set of stations.
Their architecture uses a similar CNN based feature extraction as TEAM-LM.
In contrast to TEAM-LM, for feature combination it uses maximum pooling to aggregate the feature vectors from all stations instead of a transformer.
In addition they do not add predefined position embeddings, but concatenate the feature vector for each station with the location coordinates and apply a multi-layer perceptron to get the final feature vectors for each station.
The model of \cite{vandenendeAutomatedSeismicSource2020a} is both trained and evaluated on 100~s long waveforms.
In its original form it is therefore not suitable for real time processing, although the real time processing could be added with the same zero-padding approach employed for TEAM and TEAM-LM.
The detail differences in the CNN structure and the real-time processing capability make a comparison of the exact model of \cite{vandenendeAutomatedSeismicSource2020a} to TEAM-LM difficult.

To still compare TEAM-LM to the techniques introduced in this approach, we implemented a model based on the key concepts of \cite{vandenendeAutomatedSeismicSource2020a}.
As we aim to evaluate the performance differences from the conceptual changes, rather than different hyperparameters, e.g., the exact size and number of the convolutional layers, we use the same architecture as TEAM-LM for the feature extraction and the mixture density output.
Additionally we train the model for real time processing using zero padding.
In comparison to TEAM-LM we replace the transformer with a maximum pooling operation and remove the event token.

We evaluate two different representations for the position encoding.
In the first, we concatenated the positions to the feature vectors as proposed by \cite{vandenendeAutomatedSeismicSource2020a}.
In the second, we add the position embeddings element-wise to the feature vectors as for TEAM-LM.
In both cases, we run a three-layer perceptron over the combined feature and position vector for each station, before applying the pooling operation.

We use the fast magnitude estimation approach \citep{kuyukGlobalApproachProvide2013} as a classical, i.e., non deep-learning, baseline for magnitude.
The magnitude is estimated from the horizontal peak displacement in the first seconds of the P wave.
As this approach needs to know the hypocentral distance, it requires knowledge of the event location.
We simply provide the method with the catalog hypocenter.
While this would not be possible in real time, and therefore gives the method an unfair advantage over the deep learning approaches, it allows us to focus on the magnitude estimation capabilities.
Furthermore, in particular for Italy and Japan, the high station density usually allows for sufficiently well constrained location estimates at early times.
For a full description of this baseline, see supplement section \ref{SM-sec:baseline}.

As a classical location baseline we employ NonLinLoc \citep{lomax2000probabilistic} with the  1D velocity models from \cite{graeberThreedimensionalModelsWave1999} (Chile), \cite{ueno2002improvement} (Japan) and \cite{matrullo2013improved} (Italy).
For the earliest times after the event detection usually only few picks picks are available.
Therefore we apply two heuristics.
Until at least 3/5/5 (Chile/Japan/Italy) picks are available, the epicenter is estimated as the arithmetic mean of the stations with picked arrivals so far, while the depth is set to the median depth in the training data set.
Until at least 4/7/7 picks are available, we apply NonLinLoc, but fix the depth to the median depth in the data set.
We require higher numbers of picks for Italy and Japan, as the pick quality is lower than in Chile but the station density is higher.
This leads to worse early NonLinLoc estimates in Italy and Japan compared to Chile, but improves the performance of the heuristics.

\section{Results}

\subsection{Magnitude estimation performance}

\begin{figure}
 \centering
 \includegraphics[width=\textwidth]{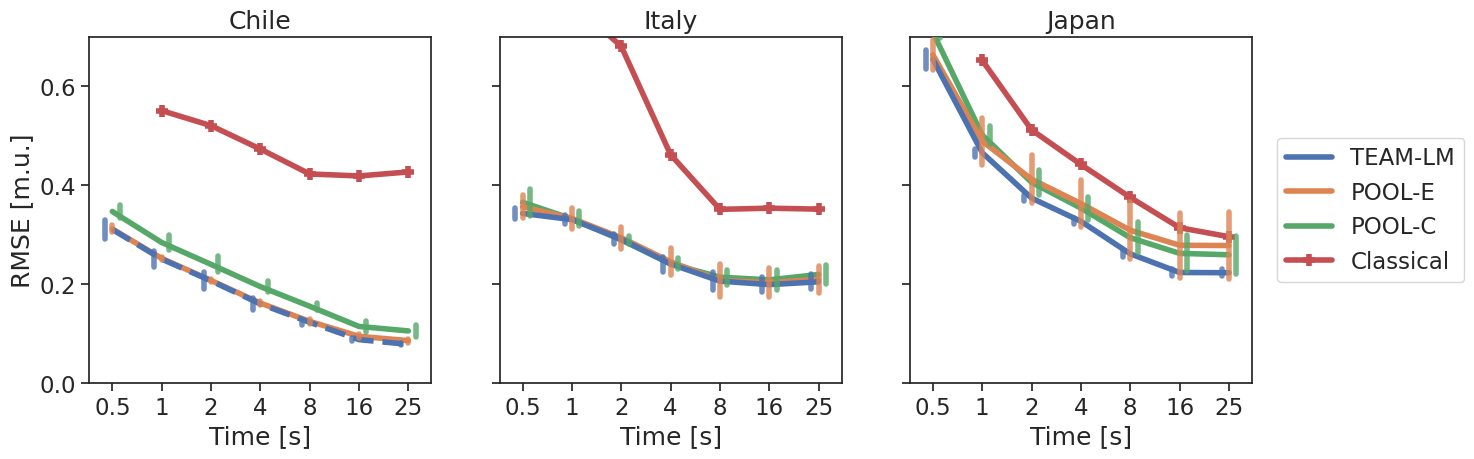}
 \caption{RMSE of the mean magnitude predictions from TEAM-LM, the pooling model with sinusoidal location embeddings (POOL-E), the pooling model with concatenated positions (POOL-C) and the classical baseline method. The time indicates the time since the first P arrival at any station, the RMSE is provided in magnitude units [m.u.]. Error bars indicate $\pm 1$ standard deviation when training the model with different random initializations. For better visibility error bars are provided with a small x-offset. Standard deviations were obtained from six realisations. Note that the uncertainty of the provided means is by a factor $\sqrt{6}$ smaller than the given standard deviation, due to the number of samples. We provide no standard deviation for the baseline, as it does not depend on a model initialization.}
 \label{fig:magnitude_rmse}
\end{figure}

\begin{figure}
 \includegraphics[width=\textwidth]{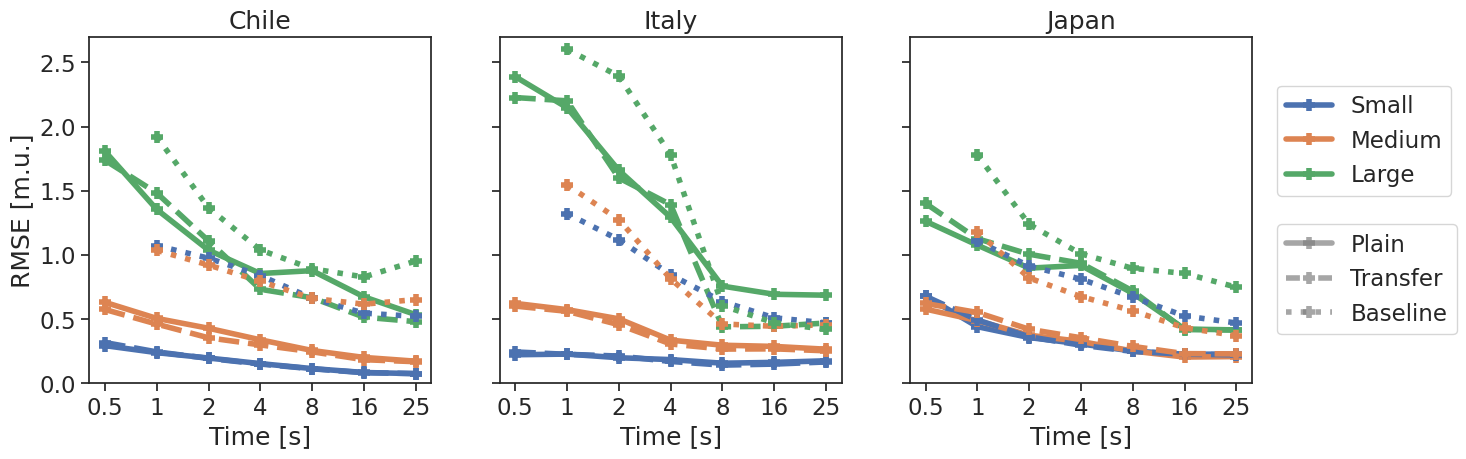}
\caption{RMSE comparison of the TEAM-LM mean magnitude predictions for different magnitude buckets. Linestyles indicate the model type: trained only on the target data (solid line), using transfer learning (dashed), classical baseline (dotted). For Chile/Italy/Japan we count events as small if their magnitude is below $3.5$/$3.5$/$4$ and as large if their magnitude is at least $5.5$/$5$/$6$. The time indicates the time since the first P arrival at any station, the RMSE is provided in magnitude units [m.u.].}
 \label{fig:magnitude_rmse_buckets}
\end{figure}

\begin{figure}
 \centering
 \includegraphics[width=\textwidth]{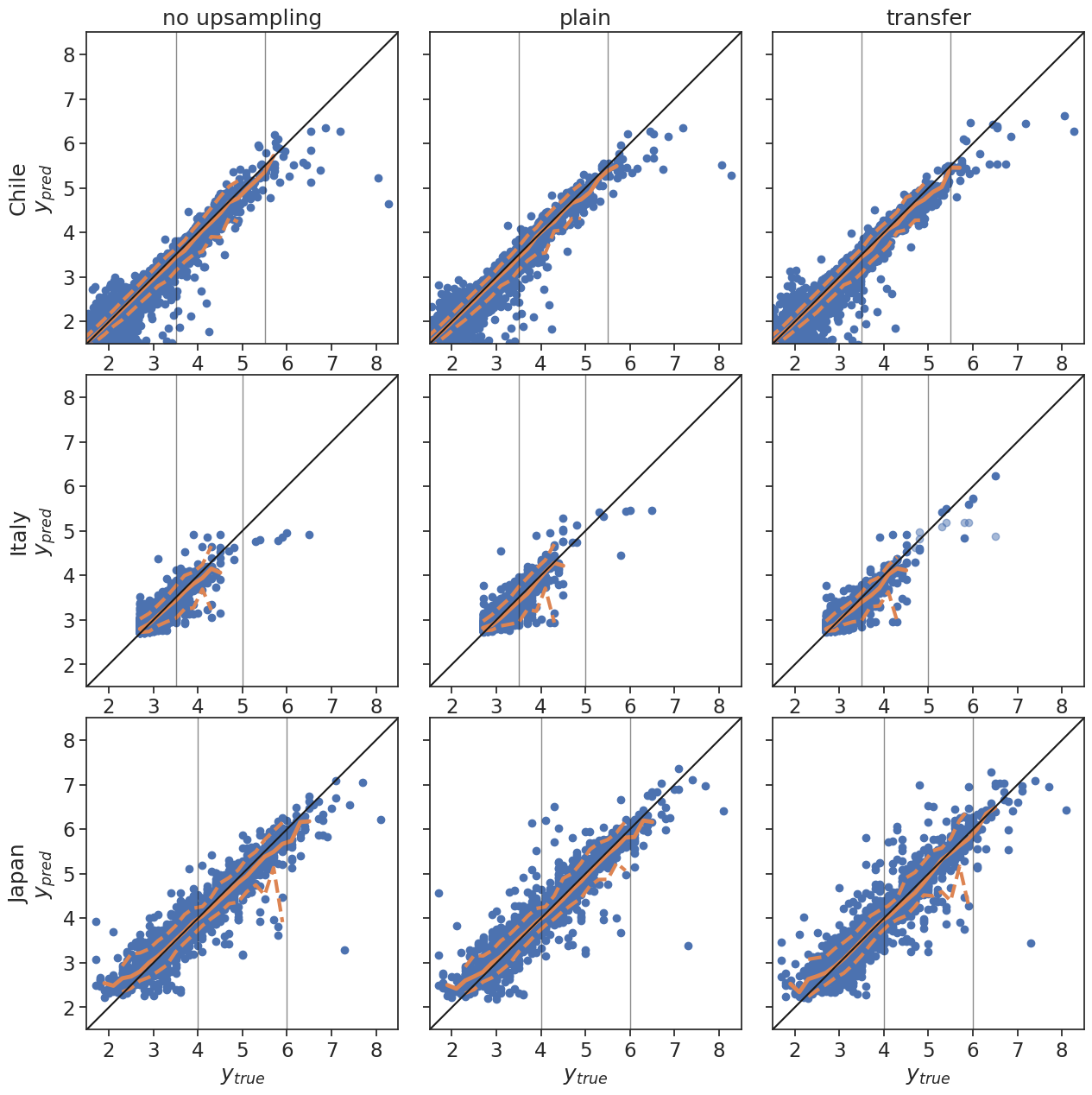}
 \caption{True and predicted magnitudes without upsampling or transfer learning (left column), with upsampling but without transfer learning (middle column) and with upsampling and transfer learning (right column). All plots show predictions after 8 seconds. In the transfer column for Chile and Japan we show results after fine-tuning on the target data set; for Italy we show results from the model without fine-tuning as this model performed better. For the largest events in Italy ($M>4.5$) we additionally show the results after fine-tuning with pale blue dots. We suspect the degraded performance in the fine tuned model results from the fact, that the largest training event ($M_W=6.1$) is considerably smaller than the largest test event ($M_W=6.5$). Vertical lines indicate the borders between small, medium and large events as defined in Figure \ref{fig:magnitude_rmse_buckets}. The orange lines show the running 5th, 50th and 95th percentile in 0.2 m.u. buckets. Percentile lines are only shown if sufficiently many data points are available. The very strong outlier for Japan (true $\sim$7.3, predicted $\sim$3.3) is an event far offshore ($>$2000~km).}
 \label{fig:transfer_magnitude}
\end{figure}

We first compare the estimation capabilities of TEAM-LM to the baselines in terms of magnitude (Figure \ref{fig:magnitude_rmse}).
We evaluate the models at fixed times $t$~=~0.5~s, 1~s, 2~s, 4~s, 8~s, 16~s, 25~s after the first P arrival at any station in the network.
In addition to presenting selected results here, we provide tables with the results of further experiments in the supplementary material (Tables \ref{SM-tab:rmse_stats}--\ref{SM-tab:mae_stats_epi}).

TEAM-LM outperforms the classical magnitude baseline consistently.
On two data sets, Chile and Italy, the performance of TEAM-LM with only 0.5~s of data is superior to the baseline with 25~s of data.
Even on the third data set, Japan, TEAM-LM requires only approximately a quarter of the time to reach the same precision as the classical baseline and achieves significantly higher precision after 25~s.
The RMSE for TEAM-LM stabilizes after 16~s for all data sets with final values of 0.08~m.u.\ for Chile, 0.20~m.u.\ for Italy and 0.22~m.u.\ for Japan.
The performance differences between TEAM-LM and the classical baseline result from the simplified modelling assumptions for the baseline.
While the relationship between early peak displacement and magnitude only holds approximately, TEAM-LM can extract more nuanced features from the waveform.
In addition, the relationship for the baseline was originally calibrated for a moment magnitude scale.
While all magnitude scales have an approximate 1:1 relationship with moment magnitude, this might introduce further errors.

We further note that the performance of the classical baseline for Italy are consistent with the results reported by \cite{festaPerformanceEarthquakeEarly2018a}.
They analyzed early warning performance in a slightly different setting, looking only at the 9 largest events in the 2016 Central Italy sequence.
However, they report a RMSE of 0.28~m.u. for the PRESTO system 4~s after the first alert, which matches approximately the 8~s value in our analysis.
Similarly, \cite{leytonHowFastCan2018} analyze how fast magnitudes can be estimated in subductions zones, and obtain values of $0.01 \pm 0.28$ across all events and $-0.70 \pm 0.30$ for the largest events ($M_w > 7.5$) at 30~s after origin time.
This matches the observed performance of the classical baseline for Japan.
For Chile, our classical baseline performs considerably worse, likely caused by the many small events with bad SNR compared to the event set considered by \citet{leytonHowFastCan2018}.
However, TEAM-LM still outperforms the performance numbers reported by \cite{leytonHowFastCan2018} by a factor of more than 2.

Improvements for TEAM-LM in comparison to the deep learning baseline variants are much smaller than to the classical approach.
Still, for the Japan data set at late times, TEAM-LM offers improvements of up to 27~\% for magnitude.
For the Italy data set, the baseline variants are on par with TEAM-LM. 
For Chile, only the baseline with position embeddings is on par with TEAM-LM.
Notably, for the Italy and Japan data sets, the standard deviation between multiple runs with different random model initialization is considerably higher for the baselines than for TEAM-LM (Figure \ref{fig:magnitude_rmse}, error bars).
This indicates that the training of TEAM-LM is more stable with regard to model initialization.

The gains of TEAM-LM can be attributed to two differences: the transformer for station aggregation and the position embeddings.
In our experiments we ruled out further differences, e.g. size and structure of the feature extraction CNN, by using identical network architectures for all parts except the feature combination across stations.
Regarding the impact of position embeddings, the results do not show a consistent pattern.
Gains for Chile seem to be solely caused by the position embeddings; gains for Italy are generally lowest, but again the model with position embeddings performs better; for Japan the concatenation model performs slightly better, although the variance in the predictions makes the differences non-significant.
We suspect these different patterns to be caused by the different catalog and network sizes as well as the station spacing.

We think that gains from using a transformer can be explained with its attention mechanism.
The attention allows the transformer to focus on specific stations, for example the stations which have recorded the longest waveforms so far.
In contrast, the maximum pooling operation is less flexible.
We suspect that the high gains for Japan result from the wide spatial distribution of seismicity and therefore very variable station distribution.
While in Italy most events are in Central Italy and in Chile the number of stations are limited, the seismicity in Japan occurs along the whole subduction zone with additional onshore events.
This complexity can likely be handled better with the flexibility of the transformer than using a pooling operation.
This indicates that the gains from using a transformer compared to pooling with position embeddings are likely modest for small sets of stations, and highest for large heterogeneous networks.

In many use cases, the performance of magnitude estimation algorithms for large magnitude events is of particular importance.
In Figure \ref{fig:magnitude_rmse_buckets} we compare the RMSE of TEAM-LM and the classical baselines binned by catalog magnitude into small, medium and large events.
For Chile/Italy/Japan we count events as small if their magnitude is below $3.5$/$3.5$/$4$ and as large if their magnitude is at least $5.5$/$5$/$6$.
We observe a clear dependence on the event magnitude.
For all data sets the RMSE for large events is higher than for intermediate sized events, which is again higher than for small events.
On the other hand the decrease in RMSE over time is strongest for larger events.
This general pattern can also be observed for the classical baseline, even though the difference in RMSE between magnitude buckets is smaller.
As both variants of the deep learning baseline show very similar trends to TEAM-LM, we omit them from this discussion.

We discuss two possible causes for these effects: (i) the magnitude distribution in the training set restricts the quality of the model optimization, (ii) inherent characteristics of large events.
Cause (i) arise from the Gutenberg-Richter distribution of magnitudes.
As large magnitudes are rare, the model has significantly less examples to learn from for large magnitudes than for small ones.
This should impact the deep learning models the strongest, due to their high number of parameters.
Cause (ii) has a geophysical origin.
As large events have longer rupture durations, the information gain from longer waveform recordings is larger for large events.
At which point during the rupture the final rupture size can be accurately predicted is a point of open discussion (e.g., \citet{meierHiddenSimplicitySubduction2017}, \citet{colombelliEarlyRuptureSignals2020}).
We probe the likely individual contributions of these causes in the following.

Estimations for large events not only show lower precision, but are also biased (Figure \ref{fig:transfer_magnitude}, middle column).
For Chile and Italy a clear saturation sets in for large events.
Interestingly the saturation starts at different magnitudes, which are around 5.5 for Italy and 6.0 for Chile.
For Japan, events up to magnitude 7 are predicted without obvious bias.
This saturation behavior is not only visible for TEAM-LM, but has also been observed in prior studies, e.g., in \citet{mousaviMachineLearningApproachEarthquake2020} (their Fig. 3, 4).
In their work, with a network trained on significantly smaller events, the saturation already set in around magnitude 3.
The different saturation thresholds indicate that the primary cause for saturation is not the longer rupture duration of large events or other inherent event properties, as in cause (ii), but is instead likely related to the low number of training examples for large events, rendering it nearly impossible to learn their general characteristics, as in cause (i).
This explanation is consistent with the much higher saturation threshold for the Japanese data set, where the training data set contains a comparably large number of large events, encompassing the year 2011 with the Tohoku event and its aftershocks.

As a further check of cause (i), we trained models without upsampling large magnitude events during training, thereby reducing the occurrence of large magnitude events to the natural distribution observed in the catalog (Figure \ref{fig:transfer_magnitude}, left column).
While the overall performance stays similar, the performance for large events is degraded on each of the data sets.
Large events are on average underestimated even more strongly.
We tried different upsampling rates, but were not able to achieve significantly better performance for large events than the configuration of the preferred model presented in the paper. 
This shows that upsampling yields improvements, but can not solve the issue completely, as it does not introduce actual additional data.
On the other hand, the performance gains for large events from upsampling seem to cause no observable performance drop for smaller event.
As the magnitude distribution in most regions approximately follows a Gutenberg-Richter law with $b \approx 1$, upsampling rates similar to the ones used in this paper will likely work for other regions as well.

The expected effects of cause (ii), inherent limitations to the predictability of rupture evolutions, can be approximated with physical models.
To this end, we look at the model from \citet{trugmanPeakGroundDisplacement2019}, which suggests a weak rupture predictability, i.e., predictability after 50~\% of the rupture duration.
\citet{trugmanPeakGroundDisplacement2019} discuss the saturation of early peak displacement and the effects for magnitude predictions based on peak displacements.
Following their model, we would expect magnitude saturation at approximately magnitude 5.7 after 1~s; 6.4 after 2~s; 7.0 after 4~s; 7.4 after 8~s.
Comparing these results to Figure \ref{fig:transfer_magnitude}, the saturation for Chile and Italy clearly occurs below these thresholds, and even for Japan the saturation is slightly below the modeled threshold.
As we assumed a model with only weak rupture predictability, this makes it unlikely that the observed saturation is caused by limitations of rupture predictability.
This implies that our result does not allow any inference on rupture predictability, as the possible effects of rupture predictability are masked by the data sparsity effects.

\subsection{Location estimation performance}

\begin{figure}
 \centering
 \includegraphics[width=\colwidth]{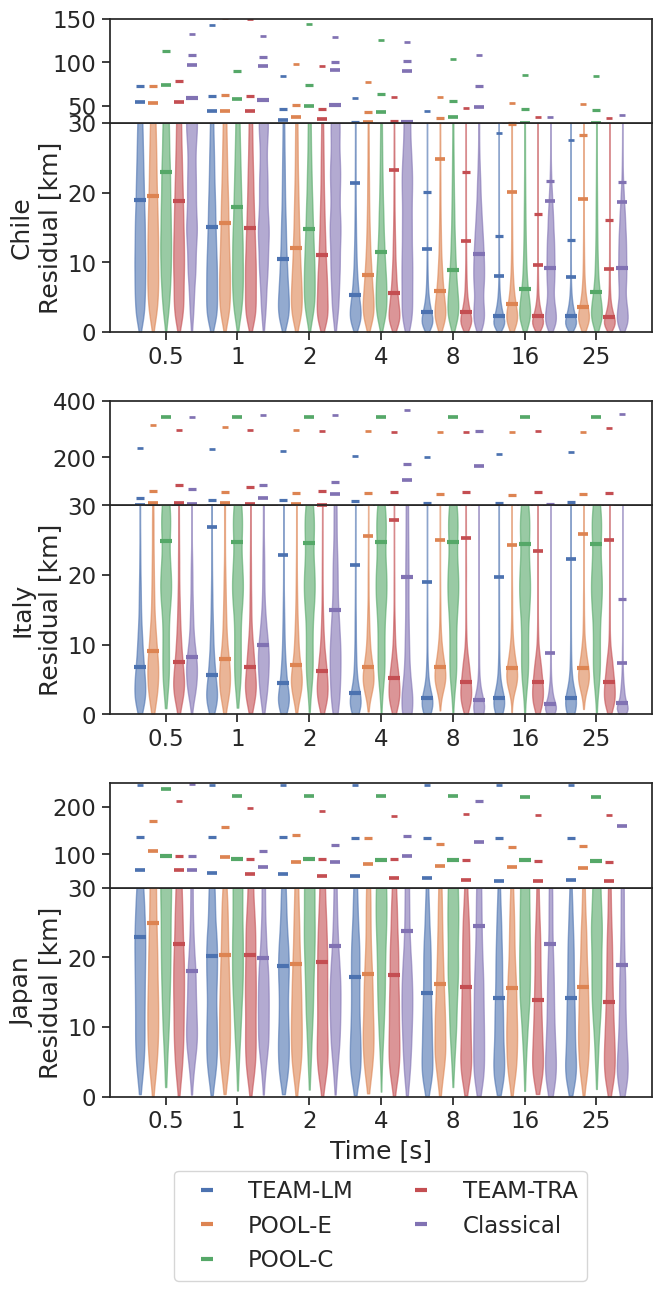}
 \caption{Violin plots and error quantiles of the distributions of the epicentral errors for TEAM-LM, the pooling baseline with position embeddings (POOL-E), the pooling baseline with concatenated position (POOL-C), TEAM-LM with transfer learning (TEAM-TRA) and a classical baseline. Vertical lines mark the 50$^{th}$, 90$^{th}$, 95$^{th}$ and 99$^{th}$ error percentiles, with smaller markers indicating higher quantiles. The time indicates the time since the first P arrival at any station. We compute errors based on the mean location predictions. A similar plot for hypocentral errors is available in the supplementary material (Figure \ref{SM-fig:loc_residuals_hypo}).}
 \label{fig:loc_residuals}
\end{figure}

We evaluate the epicentral error distributions in terms of the 50$^{th}$, 90$^{th}$, 95$^{th}$ and 99$^{th}$ error percentiles (Figure \ref{fig:loc_residuals}).
In terms of the median epicentral error, TEAM-LM outperforms all baselines in all cases, except for the classical baseline at late times in Italy.
For all data sets, TEAM-LM shows a clear decrease in median epicentral error over time.
The decrease is strongest for Chile, going from 19~km at 0.5~s to 2~km at 25~s.
For Italy the decrease is from 7~km to 2~km, for Japan from 22~km to 14~km.
For all data sets the error distributions are heavy tailed.
While for Chile even the errors at high quantiles decrease considerably over time, these quantiles stay nearly constant for Italy and Japan.

Similar to the difficulties for large magnitudes, the characteristics of the location estimation point to insufficient training data as source of errors.
The Chile data set covers the smallest region and has by far the lowest magnitude of completeness, leading to the highest event density.
Consequently the location estimation performance is best and outliers are very rare.
For the Italy and Japan data sets, significantly more events occurred in regions with only few training events, causing strong outliers.
The errors for the Japanese data set are highest, presumably related to the large number of offshore events with consequently poor azimuthal coverage.

We expect a further difference from the number of unique stations.
While for a small number of unique stations, as in the Chile data set, the network can mostly learn to identify the stations using their position embeddings, it might be unable to do so for a larger number of stations with fewer training examples per station.
Therefore the task is significantly more complicated for Italy and Japan, where the concept of station locations has to be learned simultaneously to the localization task.
This holds true even though we encode the station locations using continuously varying position embeddings.
Furthermore, whereas for  moderate and large events waveforms from all stations of the Chilean network will show the earthquake and can contribute information, the limitation to 25 stations of the current TEAM-LM implementation does not allow a full exploitation of the information contained in the hundreds of  recordings of larger events in the Japanese and Italian data sets.
This will matter in particular for out-of-network events, where the wavefront curvature and thus event distance can only be estimated properly by considering stations with later arrivals.

Looking at the classical baseline, we see that it performs considerably worse than TEAM-LM in the Chile data set in all location quantiles, better than TEAM-LM in all but the highest quantiles at late times in the Italy data set, and worse than TEAM-LM at late times in the Japan data set.
This strongly different behavior can largely be explained with the pick quality and the station density in the different data sets.
While the Chile data set contains high quality automatic picks, obtained using the MPX picker \citep{aldersons2004toward}, the Italy data set uses a simple STA/LTA and the Japan data set uses triggers from KiKNet.
This reduces location quality for Italy and Japan, in particular in the case of a low number of picks available for location.
On the other hand, the very good median performance of the classical approach for Italy can be explained from the very high station density, giving a strong prior on the location.
An epicentral error of around 2~km after 8~s is furthermore consistent with the results from \cite{festaPerformanceEarthquakeEarly2018a}.
Considering the reduction in error due to the high station density in Italy, we note that the wide station spacing in Chile likely caused higher location errors than would be achievable with a denser seismic network designed for early warning.

In addition to the pick quality, the assumption of a 1D velocity model for NonLinLoc introduces a systematic error into the localization, in particular for the subduction regions in Japan and Chile where the 3D structure deviates considerably from the 1D model.
Because of these limitations the classical baseline could be improved by employing more proficient pickers or fine-tuned velocity models.
Nonetheless, in particular the results from Chile, where the classical baseline has access to high quality P-picks, suggest that TEAM-LM can, given sufficient training data, outperform classical real-time localization algorithms. 

For magnitude estimation no consistent performance differences between the baseline approach with position embeddings and the approach with concatenated coordinates, as originally proposed by \cite{vandenendeAutomatedSeismicSource2020a}, are visible.
In contrast, for location estimation, the approach with embeddings consistently outperforms the approach with concatenated coordinates.
The absolute performance gains between the baseline with concatenation and the baseline with embeddings is even higher than the gains from adding the transformer to the embedding model.
We speculate that the positional embeddings might show better performance because they explicity encode information on how to interpolate between locations at different scales, enabling an improved exploitation of the information from stations with few or no training examples.
This is more important for location estimation, where an explicit notion of relative position is required.
In contrast, magnitude estimation can use further information, like frequency content, which is less position dependent.

\subsection{Transfer learning}

A common strategy for mitigating data sparsity is the injection of additional information from related data sets through transfer learning \citep{pan2009survey}, in our use case waveforms from other source regions. This way the model is supposed to be taught the properties of earthquakes that are consistent across regions, e.g., attenuation due to geometric spreading or the magnitude dependence of source spectra.
Note that a similar knowledge transfer implicitly is part of the classical baseline, as it was calibrated using records from multiple regions.

Here, we conduct a transfer learning experiment inspired by the transfer learning used for TEAM.
We first train a model jointly on all data sets and then fine-tune it to each of the target data sets.
This way, the model has more training examples, which is of special relevance for the rare large events, but still is adapted specifically to the target data set.
As the Japan and Italy data sets contain acceleration traces, while the Chile data set contains velocity traces, we first integrate the Japan and Italy waveforms to obtain velocity traces.
This does not have a significant impact on the model performance, as visible in the full results tables (Tables \ref{SM-tab:rmse_stats} to \ref{SM-tab:ks_stats}).

Transfer learning reduces the saturation for large magnitudes (Figure \ref{fig:transfer_magnitude}, right column).
For Italy the saturation is even completely eliminated.
For Chile, while the largest magnitudes are still underestimated, we see a clearly lower level of underestimation than without transfer learning.
Results for Japan for the largest events show nearly no difference, which is expected as the Japan data set contains the majority of large events and therefore does not gain significant additional large training examples using transfer learning.
The positive impact of transfer learning is also reflected in the lower RMSE for large and intermediate events for Italy and Chile (Figure \ref{fig:magnitude_rmse_buckets}).
These results do not only offer a way of mitigating saturation for large events, but also represent further evidence for data sparsity as the reason for the underestimation.

We tried the same transfer learning scheme for mitigating mislocations (Figure \ref{fig:loc_residuals}).
For this experiment we shifted the coordinates of stations and events such that the data sets spatially overlap.
We note that this shifting is not expected to have any influence on the single data set performance, as the relative locations of events and stations within a data set stay unchanged and nowhere the model uses absolute locations.
The transfer learning approach is reasonable, as mislocations might result from data sparsity, similarly to the underestimation of large magnitudes.
However, none of the models shows significantly better performance than the preferred models, and in some instances performance even degrades.
We conducted additional experiments where shifts were applied separately for each event, but observed even worse performance.

We hypothesize that this behaviour indicates that the TEAM-LM localization does not primarily rely on travel time analysis, but rather employs some form of fingerprinting of earthquakes.
These fingerprints could be specific scattering patterns for certain source regions and receivers.
Note that similar fingerprints are exploited in the traditional template matching approaches (e.g. \cite{shelly2007non}).
While the travel time analysis should be mostly invariant to shifts and therefore be transferable between data sets, the fingerprinting is not invariant to shifts.
This would also explain why the transfer learning, where all training samples were already in the pretraining data set and therefore their fingerprints could be extracted, outperforms the shifting of single events, where fingerprints do not relate to earthquake locations.
Similar fingerprinting is presumably also used by other deep learning methods for location estimation, e.g., by \citet{kriegerowskiDeepConvolutionalNeural2019} or \citet{perolConvolutionalNeuralNetwork2018}, however further experiments would be required to prove this hypothesis.

\section{Discussion}

\subsection{Multi-task learning}

Another common method to improve the quality of machine learning systems in face of data sparsity is multi-task learning \citep{ruder2017overview}, i.e., having a network with multiple outputs for different objectives and training it simultaneously on all objectives.
This approach has previously been employed for seismic source characterisation \citep{lomaxInvestigationRapidEarthquake2019}, but without an empirical analysis on the specific effects of multi-task learning.

We perform an experiment, in which we train TEAM-LM to predict magnitude and location concurrently.
The feature extraction and the transformer parts are shared and only the final MLPs and the mixture density networks are specific to the task.
This method is known as hard parameter sharing.
The intuition is that the individual tasks share some similarity, e.g., in our case the correct estimation of the magnitude likely requires an assessment of the attenuation and geometric spreading of the waves and therefore some understanding of the source location.
This similarity is then expected to drive the model towards learning a solution for the problem that is more general, rather than specific to the training data. The reduced number of free parameters implied by hard parameter sharing is also expected to improve the generality of the derived model, if the remaining degrees of freedom are still sufficient to extract the relevant information from the training data for each sub-task. 

Unfortunately, we actually experience a moderate degradation of performance for either location or magnitude in any data set (Tables \ref{SM-tab:rmse_stats} to \ref{SM-tab:r2_stats_cut}) when following a multi-task learning strategy.
The RMSE of the mean epicenter estimate increases by at least one third for all times and data sets, and the RMSE for magnitude stays nearly unchanged for the Chile and Japan data sets, but increases by $\sim$20\% for the Italy data set.
Our results therefore exhibit a case of negative transfer.

While it is generally not known, under which circumstances multi-task learning shows positive or negative influence \citep{ruder2017overview}, a negative transfer usually seems to be caused by insufficiently related tasks.
In our case we suspect that while the tasks are related in a sense of the underlying physics, the training data set is large enough that similarities relevant for both tasks can be learned already from a single objective.
At the same time, the particularities of the two objectives can be learned less well.
Furthermore, we earlier discussed that both magnitude and location might not actually use travel time or attenuation based approaches, but rather frequency characteristics for magnitude and a fingerprinting scheme for location.
These approaches would be less transferable between the two tasks.
We conclude that hard parameter sharing does not improve magnitude and location estimation.
Future work is required to see if other multi-task learning schemes can be applied beneficially.

\subsection{Location outlier analysis}

\begin{figure}
 \includegraphics[width=\textwidth]{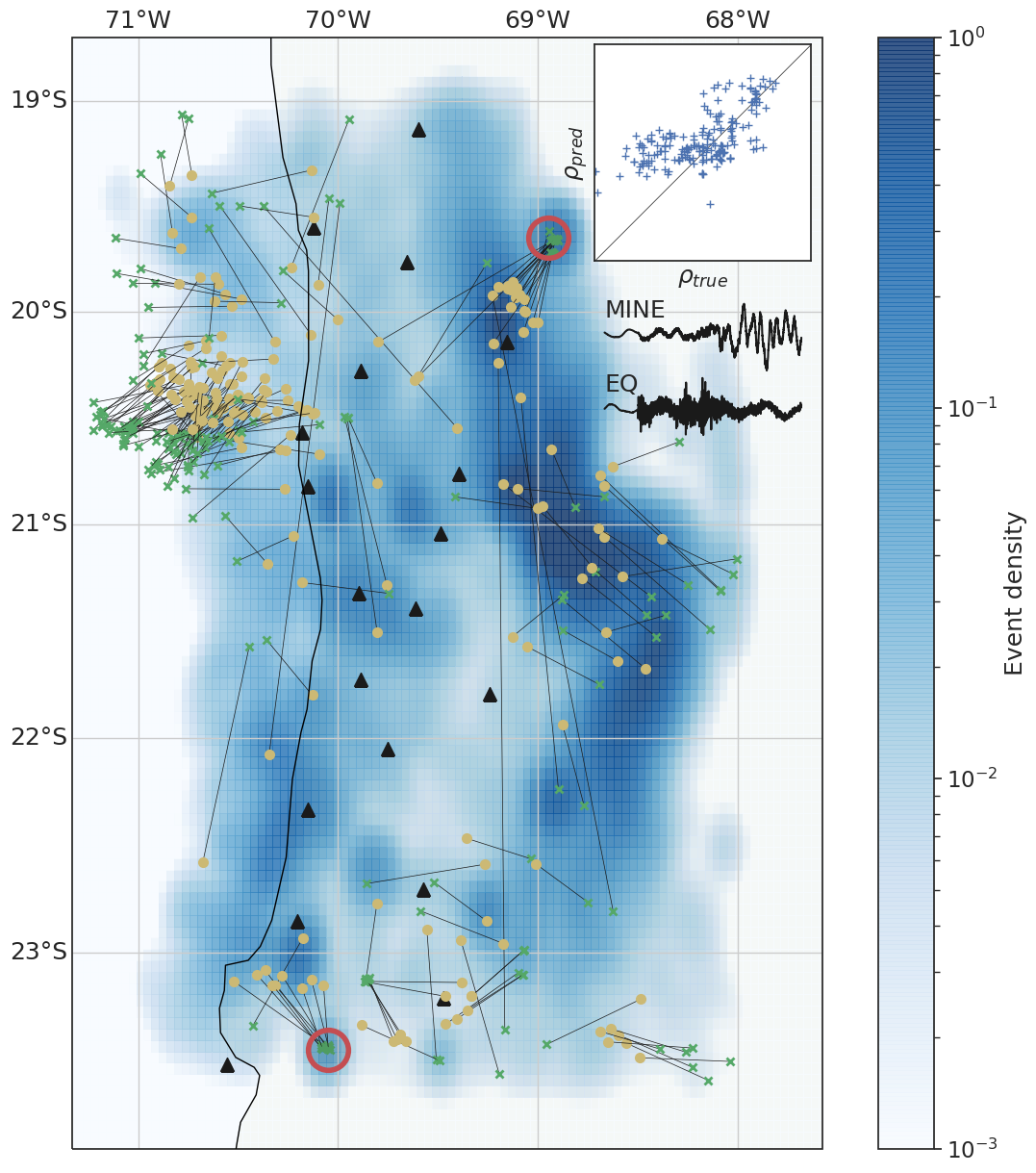}
 \caption{The 200 events with the highest location errors in the Chile data set overlayed on top of the spatial event density in the training data set. The location estimations use 16~s of data. Each event is denoted by a yellow dot for the estimated location, a green cross for the true location and a line connecting both. Stations are shown by black triangles. The event density is calculated using a Gaussian kernel density estimation and does not take into account the event depth. The inset shows the event density at the true event location in comparison to the event density at the predicted event location for the 200 events. Red circles mark locations of mine blast events. The inset waveforms show one example of a waveform from a mineblast (top) and an example waveform of an earthquake (bottom, 26~km depth) of similar magnitude ($M_A=2.5$) at similar distance (60~km) on the transverse component. Similar plots for Italy and Japan can be found in the supplementary material (Figures \ref{SM-fig:mislocations_italy} and \ref{SM-fig:mislocations_japan}).}
 \label{fig:loc_errors}
\end{figure}

As all location error distributions are heavy tailed, we visually inspect the largest deviations between predicted and catalog locations to understand the behavior of the localization mechanism of TEAM-LM.
We base this analysis on the Chile data set (Figure \ref{fig:loc_errors}), as it has generally the best location estimation performance, but observations are similar for the other data sets (Figures \ref{SM-fig:mislocations_italy} and \ref{SM-fig:mislocations_japan}).

Nearly all mislocated events are outside the seismic network and location predictions are generally biased towards the network.
This matches the expected errors for traditional localization algorithms.
In contrast to traditional algorithms, events are not only predicted to be closer to the network, but they are also predicted as lying in regions with a higher event density in the training set (Figure \ref{fig:loc_errors}, inset).
This suggests that not enough similar events were included in the training data set.
Similarly, \cite{kriegerowskiDeepConvolutionalNeural2019} observed a clustering tendency when predicting the location of swarm earthquakes with deep learning.

We investigated two subgroups of mislocated events: the Iquique sequence, consisting of the Iquique mainshock, foreshocks and aftershocks, and mine blasts.
The Iquique sequence is visible in the north-western part of the study area.
All events are predicted approximately 0.5$^\circ$ too far east.
The area is both outside the seismic network and has no events in the training set.
This systematic mislocation may pose a serious threat in applications, such as early warning, when confronted with a major change in the seismicity pattern, as is common in the wake of major earthquakes or during sudden swarm activity, which are also periods of heightened seismic hazard. 

For mine blasts, we see one mine in the northeast and one in the southwest (marked by red circles in Figure \ref{fig:loc_errors}).
While all events are located close by, the location are both systematically mispredicted in the direction of the network and exhibit scatter.
Mine-blasts show a generally lower location quality in the test set.
While they make up only $\sim$1.8\% of the test set, they make up 8\% of the top 500 mislocated events.
This is surprising as they occur not only in the test set, but also in similar quantities in the training set.
We therefore suspect that the difficulties are caused by the strongly different waveforms of mine blasts compared to earthquakes.
One waveform of each a mine blast and an earthquake, recorded at similar distances are shown as inset in Figure \ref{fig:loc_errors}.
While for the earthquake both a P and S wave are visible, the S wave can not be identified for the mine blast.
In addition, the mine blast exhibits a strong surface wave, which is not visible for the earthquake.
The algorithm therefore can not use the same features as for earthquakes to constrain the distance to a mine blast event.

\subsection{The impact of data set size and composition}
\label{sec:data_availability}

\begin{figure}
 \centering
 \includegraphics[width=0.85\textwidth]{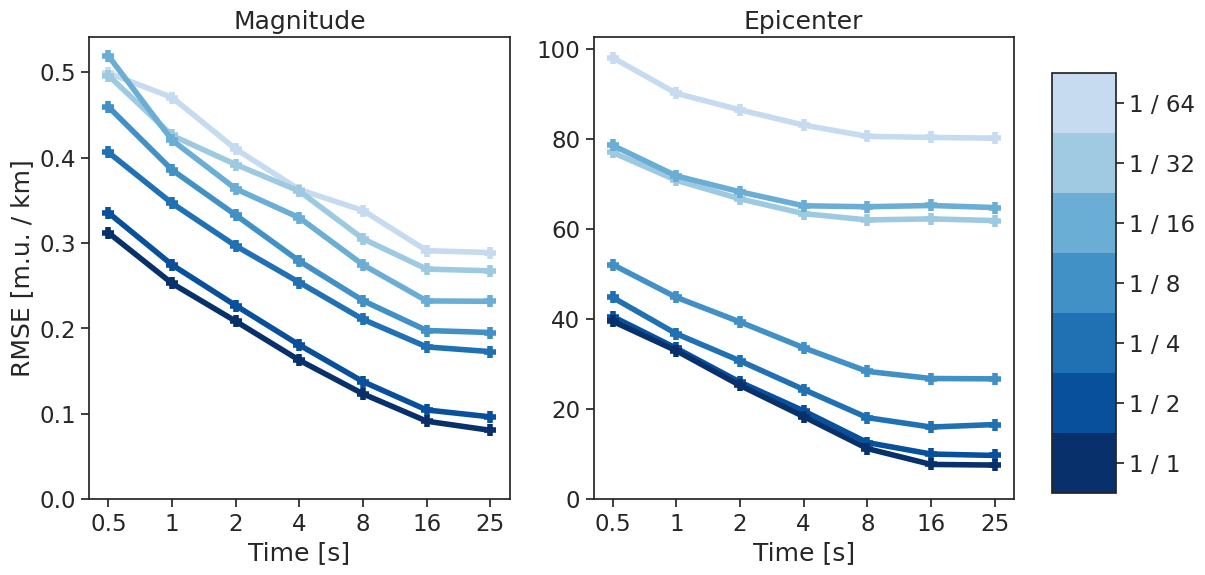}
 \caption{RMSE for magnitude and epicentral location at different times for models trained on differently sized subsets of the training set in Chile. The line color encodes the fraction of the training and validation set used in training. All models were evaluated on the full Chilean test set. We note that the variance of the curves with fewer data is higher, due to the increased stochasticity from model training and initialization.}
 \label{fig:data_sparsity_rmse}
\end{figure}

Our analysis so far showed the importance of the amount of training data.
To quantify the impact of data availability on magnitude and location estimation, we trained models only using fractions of the training and validation data (Figure \ref{fig:data_sparsity_rmse}).
We use the Chile data set for this analysis, as it contains by far the most events.
We subsample the events by only using each $k^{th}$ event in chronological order, with $k=2, 4, 8, 16, 32, 64$.
This strategy approximately maintains the magnitude and location distribution of the full set.
We point out, that TEAM-LM only uses information of the event under consideration and does not take the events before or afterwards into account.
Therefore, the `gaps' between events introduced by the subsampling do not negatively influence TEAM-LM.

For all times after the first P arrival, we see a clear increase in the magnitude-RMSE for a reduction in the number of training samples.
While the impact of reducing the data set by half is relatively small, using only a quarter of the data already leads to a twofold increase in RMSE at late times.
Even more relevant in an early warning context, a fourfold smaller data sets results in an approximately fourfold increase in the  time needed to reach the same precision as with the full data.
This relationship seems to hold approximately across all subsampled data sets: reducing the data set $k$ fold increases the time to reach a certain precision by a factor of $k$.

We make three further observations from comparing the predictions to the true values (Figure \ref{SM-fig:data_sparsity_true_pred}).
First, for nearly all models the RMSE changes only marginally between 16~s and 25~s, but the RMSE of this plateau  increases significantly with a decreasing number of training events.
Second, the lower the amount of training data, the lower is the saturation threshold above which all events are strongly underestimated.
In addition, for 1/32 and 1/64 of the full data set, an `inverse saturation' effect is noticeable for the smallest magnitudes.
Third, while for the full data set and the largest subsets all large events are estimated at approximately the saturation threshold, if at most one quarter of the training data is used, the largest events even fall significantly below the saturation threshold.
For the models trained on the smallest subsets (1/8 to 1/64), the higher the true magnitude the lower the predicted magnitude becomes.
We assume that the larger the event is, the further away from the training distribution it is and therefore it is estimated approximately at the most dense region of the training label distribution.
These observations support the hypothesis that underestimations of large magnitudes for the full data set are caused primarily by insufficient training data.

While the RMSE for epicenter estimation shows a similar behavior as the RMSE for magnitude, there are subtle differences.
If the amount of training data is halved, the performance only degrades mildly and only at later times.
However, the performance degradation is much more severe than for magnitude if only a quarter or less of the training data are available.
This demonstrates that location estimation with high accuracy requires catalogs with a high event density.

The strong degradation further suggests insights into the inner working of TEAM-LM.
Classically, localization should be a task where interpolation leads to good results, i.e., the travel times for an event in the middle of two others should be approximately the average between the travel times for the other events.
Following this argument, if the network would be able to use interpolation, it should not suffer such significant degradation when faced with fewer data.
This provides further evidence that the network does not actually learn some form of triangulation, but only an elaborate fingerprinting scheme, backing the finding from the qualitative analysis of location errors.

\subsection{Training TEAM-LM on large events only}
Often, large events are of the greatest concerns, and as discussed, generally showed poorer performance because they are not well represented in the training data.
It therefore appears plausible that a model optimized for large events might perform better than a model trained on both large and small events.
In order to test this hypothesis, we employed an extreme version of the upscaling strategy by training a set of models only on large events, which might avoid tuning the model to seemingly irrelevant small events.
In fact, these models perform significantly worse than the models trained on the full data set, even for the large events (Tables \ref{SM-tab:rmse_stats} to \ref{SM-tab:r2_stats_cut}).
Therefore even if the events of interest are only the large ones, training on more complete catalogs is still beneficial, presumably by giving the network more comprehensive information on the regional propagation characteristics and possibly site effects.

\subsection{Interpretation of predicted uncertainties}

\begin{figure}
 \centering
 \includegraphics[width=\colwidth]{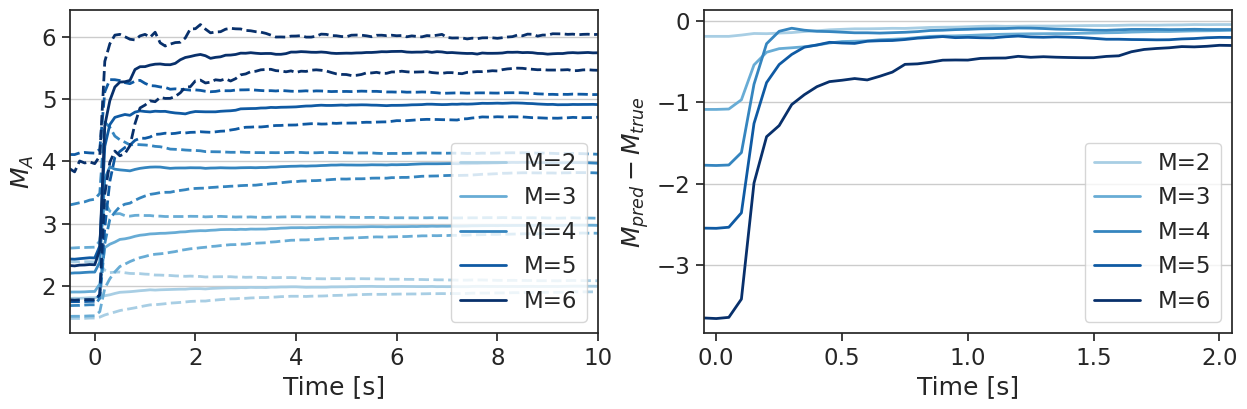}
 \caption{Magnitude predictions and uncertainties in the Chile data set as a function of time since the first P arrival. Solid lines indicate median predictions, while dashed lines (left panel only) show 20th and 80th quantiles of the prediction. The left panel shows the predictions, while the right panel shows the differences between the predicted and true magnitude. The right panel is focused on a shorter timeframe to show the early prediction development in more detail. In both plots, each color represents a different magnitude bucket. For each magnitude bucket, we sampled 1,000 events around this magnitude and combined their predictions. If less than 1,000 events were available within $\pm$0.5~m.u. of the bucket center, we use all events within this range. We only use events from the test set. To ensure that the actual uncertainty distribution is visualized, rather than the distribution of magnitudes around the bucket center, each prediction is shifted by the magnitude difference between bucket center and catalog magnitude.}
 \label{fig:mag_uncertainties_over_time}
\end{figure}

\begin{figure}
 \centering
 \includegraphics[width=\colwidth]{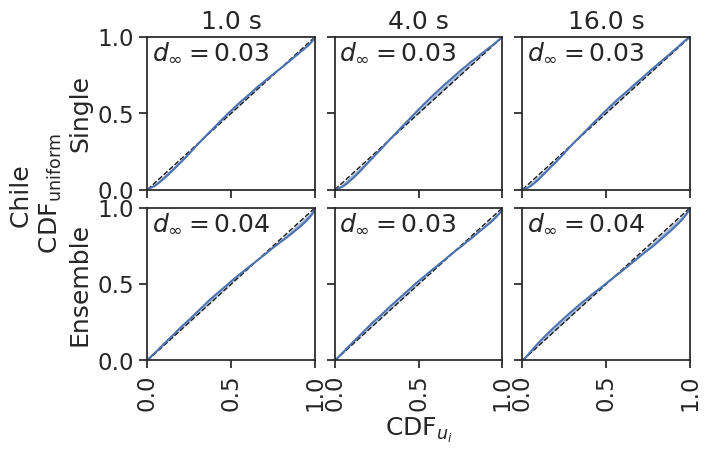}\\
 \includegraphics[width=\colwidth]{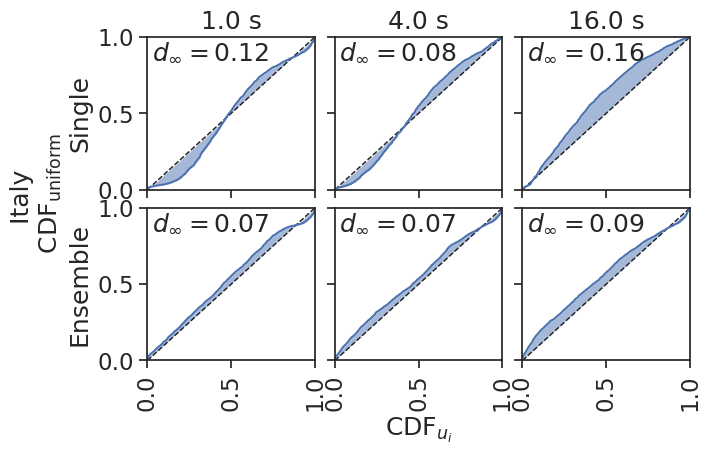}\\
 \includegraphics[width=\colwidth]{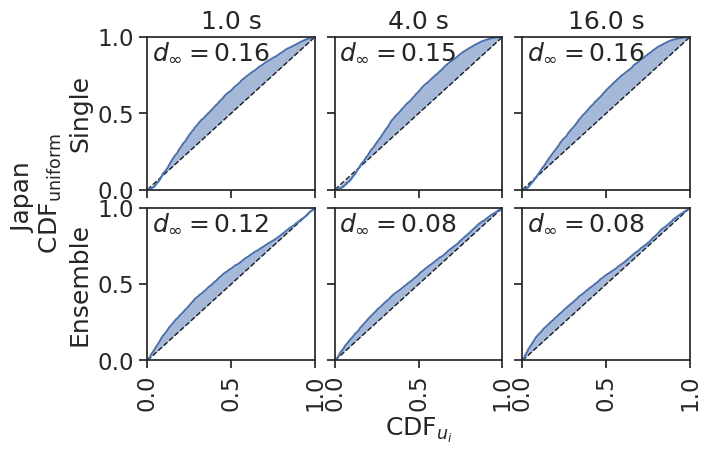}\\
 \caption{P-P plots of the CDFs of the empirical quantile of the magnitude predictions compared to the expected uniform distribution. The P-P plot shows $(\mathrm{CDF}_{u_i}(z), \mathrm{CDF}_{\mathrm{uniform}}(z))$ for $z \in [0, 1]$. The expected uniform distribution is shown as the diagonal line, the misfit is indicated as shaded area. The value in the upper corner provides $d_\infty$, the maximum distance between the diagonal and the observed CDF. $d_\infty$ can be interpreted as the test statistic for a Kolmogorov-Smirnov test. Curves consistently above the diagonal indicate a bias to underestimation, and below the diagonal to overestimation. Sigmoidal curves indicate over-confidence, mirrored sigmoids indicate under-confidence.
 See supplementary section \ref{SM-sec:calibration} for a further discussion of the plotting methodology and its connection to the Kolmogorov-Smirnov test.}
 \label{fig:calibration}
\end{figure}

\begin{figure}
 \centering
 \includegraphics[width=\textwidth]{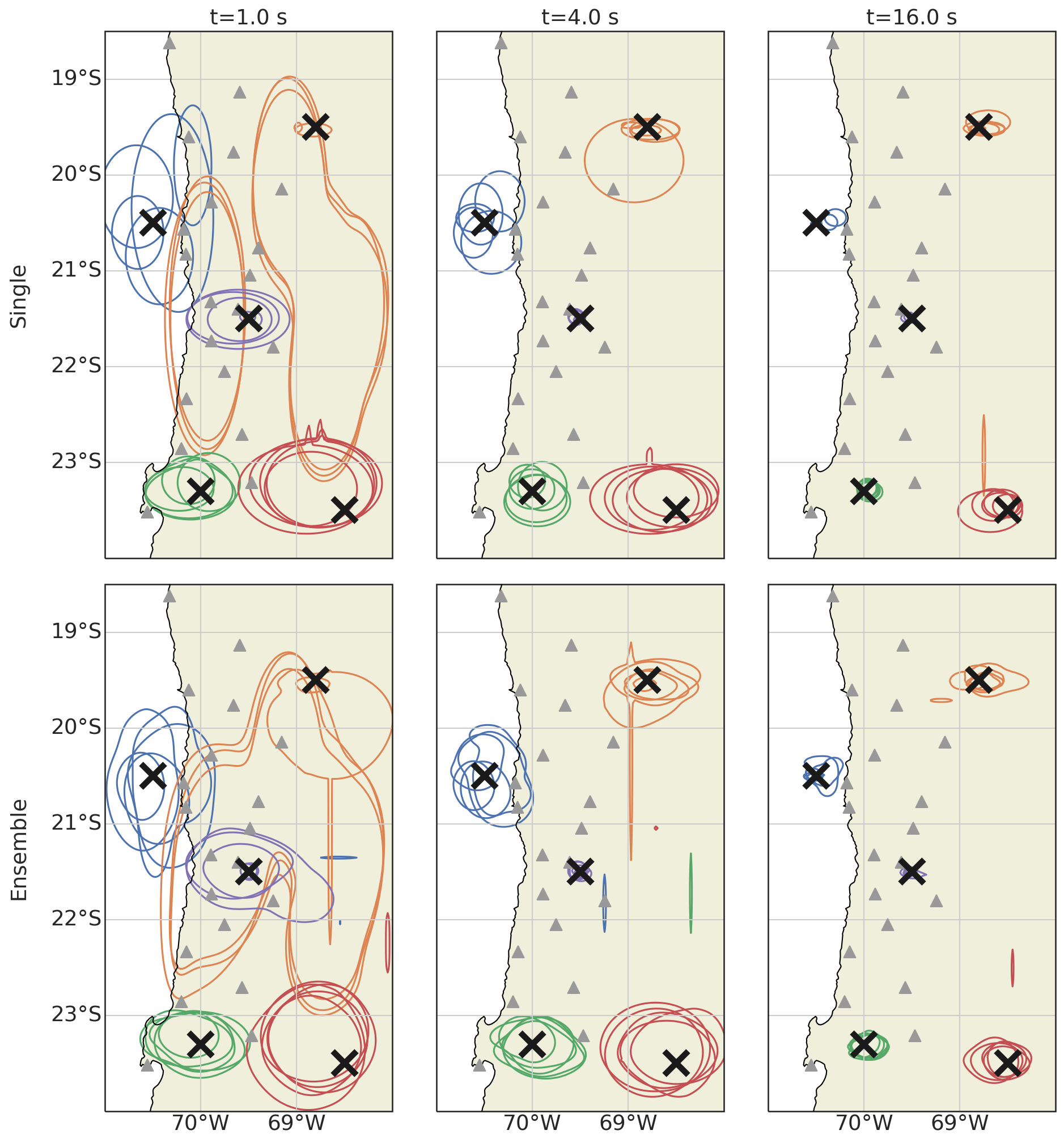}
 \caption{The figure shows 90th percent confidence areas for sample events around 5 example locations.
 For each location the 5 closest events are shown.
 Confidence areas belonging to the same location are visualized using the same color.
 Confidence areas were chosen as curves of constant likelihood, such that the probability mass above the likelihood equals 0.9.
 To visualize the result in 2D we marginalize out the depth.
 Triangles denote station locations for orientation.
 The top row plots show results from a single model, while the bottom row plots show results from an ensemble of 10 models.}
 \label{fig:location_uncertainties}
\end{figure}

So far we only analyzed the mean predictions of TEAM-LM.
As for many application scenarios, for example early warning, quantified uncertainties are required, TEAM-LM outputs not only these mean predictions, but a probability density.
Figure \ref{fig:mag_uncertainties_over_time} shows the development of magnitude uncertainties for events from different magnitude classes in the Chile data set.
The left panel shows the absolute predictions, while the right panel shows the difference between prediction and true magnitude and focuses on the first 2~s.
As we average over multiple events, each set of lines can be seen as a prototype event of a certain magnitude.

For all magnitude classes the estimation shows a sharp jump at $t=0$, followed by a slow convergence to the final magnitude estimate.
We suspect that the magnitude estimation always converges from below, as due to the Gutenberg-Richter distribution, lower magnitudes are more likely \textit{a priori}.
The uncertainties are largest directly after $t=0$ and subsequently decrease, with the highest uncertainties for the largest events.
As we do not use transfer learning in this approach, there is a consistent underestimation of the largest magnitude events, visible from the incorrect median predictions for magnitudes 5 and 6.

While the Gaussian mixture model is designed to output uncertainties, it cannot be assumed that the predicted uncertainties are indeed well calibrated, i.e., that they actually match the real error distribution.
Having well calibrated uncertainties is crucial for downstream tasks that rely on the uncertainties.
Neural networks trained with a log-likelihood loss generally tend to be overconfident \citep{snoek2019can,guo2017calibration}, i.e., underestimate the uncertainties.
This overconfidence is probably caused by the strong overparametrization of neural network models.
To assess the quality of our uncertainty estimations for magnitude, we use the observation that for a specific event $i$, the predicted Gaussian mixture implies a cumulative distribution function $F^i_{pred}:\mathbb{R} \rightarrow [0,1]$.
Given the observed magnitude $y^i_{true}$, we can calculate $u_i = F^i_{pred}(y^i_{true})$.
If $y^i_{true}$ is indeed distributed according to $F^{i}_{pred}$, then $u_i$ needs to be uniformly distributed on $[0,1]$.
We test this based on the $u_i$ of all events in the test set using P-P plots (Figure \ref{fig:calibration}).
Further details on the method can be found in the supplementary material (Section \ref{SM-sec:calibration}).
Note that good calibration is a necessary but not sufficient condition for a good probabilistic forecast.
An example of a perfectly calibrated but mostly useless probabilistic prediction would be the marginal probability of the labels.

Figure \ref{fig:calibration} shows the P-P plots of $u$ in comparison to a uniform distribution.
For all data sets and all times the model is signficantly miscalibrated, as estimated using Kolmgorov-Smirnov test statistics (Section \ref{SM-sec:calibration}).
Miscalibration is considerably stronger for Italy and Japan than for Chile.
More precisely, the model is always overconfident, i.e., estimates narrower confidence bands than the actually observed errors.
Further, in particular at later times, the model is biased towards underestimating the magnitudes.
This is least visible for Chile.
We speculate that this is a result of the large training data set for Chile, which ensures that for most events the density of training events in their magnitude range is high.

To mitigate the miscalibration, we trained ensembles \citep{hansen1990neural}, a classical method to improve calibration.
Instead of training a single neural network, a set of $n$ neural networks, in our case $n=10$, are trained, which all have the same structure, but different initialization and batching in training.
The networks therefore represent a sample of size $n$ from the posterior distribution of the model parameters given the training data.
For Italy and Japan, this improves calibration considerably (Figure \ref{fig:calibration}).
For Chile, the ensemble model, in contrast to the single model, exhibits underconfidence, i.e., estimates too broad uncertainty bands.

The maximum distance between the empirical cumulative distribution function of $u$ and a uniformly distributed variable $d_\infty$ is the test statistic of the Kolmogorov-Smirnov test.
While $d_\infty$ is reduced by nearly half for some of the ensemble results, the Kolmogorov-Smirnov test indicates, that even the distributions from the ensemble models deviate highly significantly from a uniform distribution ($p \ll 10^{-5}$).
A table with $d_\infty$ for all experiments can be found in the supplementary material (Table \ref{SM-tab:ks_stats}).

To evaluate the location uncertainties qualitatively, we plot confidence ellipses for a set of events in Chile (Figure \ref{fig:location_uncertainties}).
Again we compare the predictions from a single model to the predictions of an ensemble.
At early times, the uncertainty regions mirror the seismicity around the station with the first arrival, showing that the model correctly learned the prior distribution.
Uncertainty ellipses at late times approximately match the expected uncertainty ellipses for classical methods, i.e., they are small and fairly round for events inside the seismic network, where there is good azimuthal coverage, and larger and elliptical for events outside the network.
Location uncertainties are not symmetric around the mean prediction, but show higher likelihood towards the network than further outwards.
Location errors for the ensemble model are more smooth than from the single model, but show the same features.
The uncertainty ellipses are slightly larger, suggesting that the single model is again overconfident.

In addition to improving calibration, ensembles also lead to slight improvements regarding the accuracy of the mean predictions (Tables SM 5 to SM 11).
Improvements in terms of magnitude RMSE range up to $\sim 10\%$, for epicentral location error up to $\sim 20\%$.
Due to the high computational demand of training ensembles, all other results reported in this paper are calculated without ensembling.
We note that in addition to ensembles a variety of methods have been developed to improve calibration or obtain calibrated uncertainties.
For a quantitative survey, see for example \citet{snoek2019can}.
One of these methods, Monte-Carlo Dropout, has already been employed in the context of fast assessment by \citet{vandenendeAutomatedSeismicSource2020a}.

\section{Conclusion}

In this study we adapted TEAM to build TEAM-LM, a real time earthquake source characterization model, and used it to study the pitfalls and particularities of deep learning for this task.
We showed that TEAM-LM achieves state of the art in magnitude estimation, outperforming both a classical baseline and a deep learning baseline.
Given sufficiently large catalogs, magnitude can be assessed with a standard deviation of $\sim$0.2 magnitude units within 2~s of the first P arrival and a standard deviation of  0.07~m.u. within the first 25~s.
For location estimation, TEAM-LM outperforms a state of the art deep learning baseline and compares favorably with a classical baseline.

Our analysis showed that the quality of model predictions depends crucially on the training data.
While performance in regions with abundant data is excellent, in regions of data sparsity, prediction quality degrades significantly.
For magnitude estimation this effect results in the underestimation of large magnitude events; for location estimation events in regions with few or no training events tend to be mislocated most severely.
This results in a heavy tailed error distribution for location estimation.
Large deviations in both magnitude and location estimation can have significant impact in application scenarios, e.g., for early warning where large magnitudes are of the biggest interest.

Following our analysis, we propose a set of best practices for building models for fast earthquake source characterisation:
\begin{enumerate}
 \item Build a comprehensive evaluation platform. Put a special focus on outliers and rare or large events. Analyze which impact outliers or out of distribution events will have for the proposed application.
 \item Use very large training catalogs, spanning long time spans and having a low magnitude of completeness. If possible, employ transfer learning. We hope the catalogs used in this study can give a starting point for transfer learning.
 \item Use training data augmentation, especially upsampling of large events, which improves prediction performance in face of label sparsity at virtually no cost.
 \item If probabilistic estimates are required, use deep ensembles to improve the model calibration.
 \item When using deep learning for location estimation, put special emphasis on monitoring possible distribution shifts between training data and application.
\end{enumerate}

While these points give guidance for training current models they also point to  further directions for methodological advances.
First, our transfer learning scheme is fairly simple.
More refined and targeted schemes could increase the amount of information sharable across data sets considerably.
We further expect major improvements from training with simulated data, but are aware that generating realistic, synthetic seismogramms, especially for large events, poses major challenges.
Another promising alternative might be to move away from the paradigm of black box modeling, i.e., training algorithms that are built solely by fitting recorded data.
Instead, incorporation of physical knowledge and a move towards physics informed deep learning methods seems promising \citep{raissi2019physics}.
However, physics informed neural networks are still in their infancy and the application to seismic tasks still needs to be developed.

\section*{Data availability}
The Italy data set has been published as \citet{munchmeyer2020dataitaly}.
The Chile data set has been published as \citet{munchmeyer2021datachile}.
An implementation of TEAM-LM and TEAM has been published as \citet{munchmeyer2021team}.
Download instructions for the Japan data set are available in the code publication.

\section*{Acknowledgments}
We thank the National Research Institute for Earth Science and Disaster Resilience for providing the catalog and waveform data for our Japan data set.
We thank the Istituto Nazionale di Geofisica e Vulcanologia and the Dipartimento della Protezione Civile for providing the catalog and waveform data for our Italy data set.
We thank Christian Sippl for providing the P picks for the Chile catalog.
We thank Sebastian Nowozin for insightful discussions on neural network calibration and probabilistic regression.
Jannes Münchmeyer acknowledges the support of the Helmholtz Einstein International Berlin Research School in Data Science (HEIBRiDS).
We thank Martijn van den Ende for his comments that helped improve the manuscript.
We use obspy \citep{beyreuther2010obspy}, tensorflow \citep{abadi2016tensorflow} and color scales from \cite{crameri2018geodynamic}.

\bibliographystyle{gji}

\clearpage
\appendix

\renewcommand\thefigure{SM~\arabic{figure}}
\renewcommand\thetable{SM~\arabic{table}}
\renewcommand\thesection{SM~\arabic{section}}
\setcounter{figure}{0} 
\setcounter{table}{0}

\section{Data sources}
\begin{table}
 \centering
 \caption{Seismic networks}
 \input{seismic_networks.tex}
 \label{tab:seismic_networks}
\end{table}

\section{Classical magnitude estimation baseline}
\label{SM-sec:baseline}

For magnitude estimation we compare TEAM-LM to a classical baseline.
To this end we use the peak displacement based approach proposed by \cite{kuyukGlobalApproachProvide2013}.
At each station, we bandpass filter the signal between 0.5~Hz and 3~Hz and discard traces with insufficient signal to noise ratio.
We extract peak displacement $PD$ from the horizontal components in the first 6~s of the P wave, while only including samples before the S onset.
We use the relationship
\begin{align}
 M = c_1 \log(PD) + c_2 \log(R) + c_3 + \mathscr{N}(0, \sigma^2)
\end{align}
from \cite{kuyukGlobalApproachProvide2013} to estimate magnitudes from peak displacement.
We use $c_1 = 1.23$, $c_2 = 1.38$ and $\sigma = 0.31$ from \cite{kuyukGlobalApproachProvide2013}.
These parameters were calibrated using data from California and Japan, but the authors state that the relationship can be applied to earthquake source zones around the world.
To account for a constant offset between different magnitude scales, we optimized $c_3$ separately for each dataset such that the predictions do not have a systematic bias compared to the ground truth.

We average the predictions from multiple stations, effectively assuming independence between the predictions.
To obtain earlier predictions, we already calculate magnitude estimates at a station once at least 1~s of P wave data has been recorded.
We assign higher weights to stations with longer P wave records, with weights linearly increasing from 0.11 for 1~s of waveforms, to 1.0 for 6~s of data.
Thereby, while getting early estimates from the first stations, new data from later stations does not perturb the prediction strongly until enough data has been recorded.

As the estimation relies on the hypocentral distance $R$ between station and event, the method requires an estimate of the hypocentral location.
We provide the method with the cataloged hypocentral location.
While this is an unrealistically optimistic assumption for an actual real time determination, it allows us to put our focus on the magnitude estimation capabilities.
We note that this gives the baseline an advantage compared to TEAM-LM, which has no information on the earthquake location.

For some events in the Chile catalog, the SNR criterion is not fulfilled at any station due to the inclusion of smaller magnitude events and the higher distances between stations and events.
For these events the baseline does not issue a magnitude estimation.
We exclude these events from the evaluation of the baseline, leading to an optimistic assessment of the performance of the baseline.

\section{Calibration estimation}
\label{SM-sec:calibration}

Calibration of a model describes whether the predicted uncertainties match the observed values, i.e., if the observation $y_{true}$ was drawn from a distribution with cumulative distribution function (CDF) $F_{pred}$.
Unfortunately, for each event $i$, only one prediction observation of the magnitude $y^i_{true}$ and one prediction $F^i_{pred}$ is available.
To this end, we define the random variable $u_i = F^i_{pred}(y^i_{true})$.
If $y^i_{true}$ is distributed according to $F^i_{pred}$ than $u_i$ must be uniformly distributed on $[0, 1]$.
This follows from the definition of the CDF.
If $F$ is a CDF and $U$ a uniform random variable on $[0,1]$, than $F^{-1}(U)$ is distributed according to $F$.

We take the $u_i$ of all events as samples of a random variable $U$ and compare $U$ to a uniform random variable on $[0, 1]$.
The maximum difference between the empirical CDF of $U$ and a uniform variable is the test statistic of a Kolmogorov-Smirnov test, $d_\infty$.
As the number of events $n$ is large, critical values $d_\alpha$ to a confidence threshold $\alpha$ can be estimated as:
\begin{align}
 d_\alpha = \frac{\sqrt{-\frac{1}{2} \log{\frac{\alpha}{2}}}}{\sqrt{n}}
\end{align}
For $\alpha=10^{-5}$, this gives values $d_\alpha$ of 0.015 (Chile), 0.054 (Japan) and 0.039 (Italy).
This is considerably below the observed values $d_\infty$, even using ensembles, indicating that $U$ differs highly significantly from a uniform distribution.

\section{Figures and tables}

\begin{table}
 \centering
 \caption{Architecture of the feature extraction network. The input shape of the waveform data is (time, channels). FC denotes fully connected layers. As FC layers can be regarded as 0D convolutions, we write the output dimensionality in the filters column. The ``Concatenate scale'' layer concatenates the log of the peak amplitude to the output of the convolutions. We want to mention that depending on the existence of borehole data the number of input filters for the first Conv1D varies.}
 \input{convolutions.tex}
 \label{SM-tab:convolutions}
\end{table}

\begin{table}
 \centering
 \caption{Architecture of the transformer network.}
 \input{transformer.tex}
 \label{SM-tab:transformer}
\end{table}

\begin{table}
 \centering
 \caption{Architecture of the mixture density network.}
 \begin{tabular}{lc}
    Feature & Value \\
    \hline
    Dimensions fully connected layers (magnitude) & 150, 100, 50, 30, 10 \\
    Dimensions fully connected layers (location) & 150, 100, 50, 50, 50 \\
    Mixture size (magnitude) & 5 \\
    Mixture size (location) & 15 \\
    Base distribution & Gaussian \\
 \end{tabular}
 \label{SM-tab:mixture}
\end{table}

\begin{figure}
 \includegraphics[width=\textwidth]{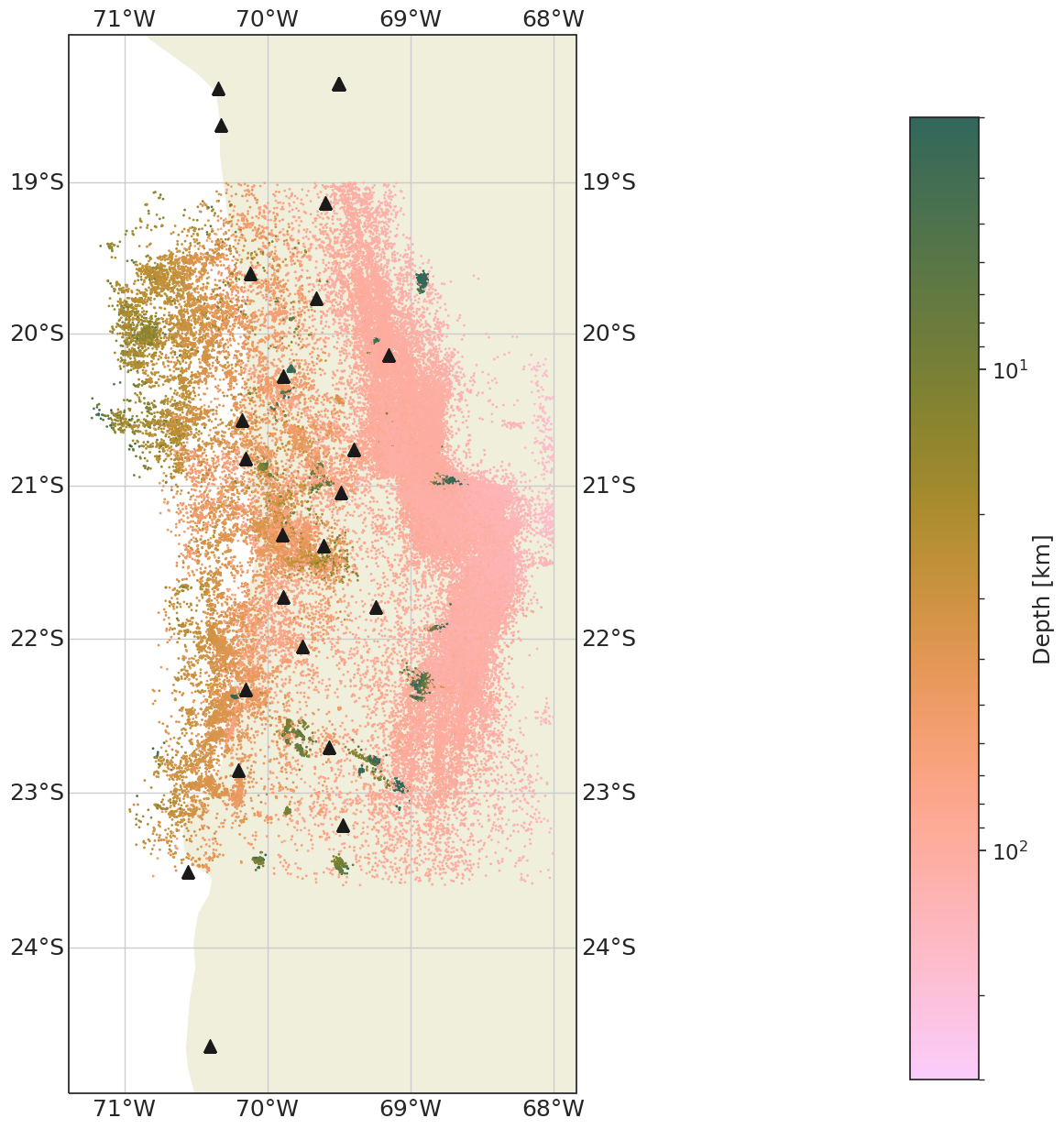}
 \caption{Event and station distribution for Chile. In the map, events are indicated by dots, stations by triangles. The event depth is encoded using color.}
 \label{SM-fig:highres_chile}
\end{figure}

\begin{figure}
 \includegraphics[width=\textwidth]{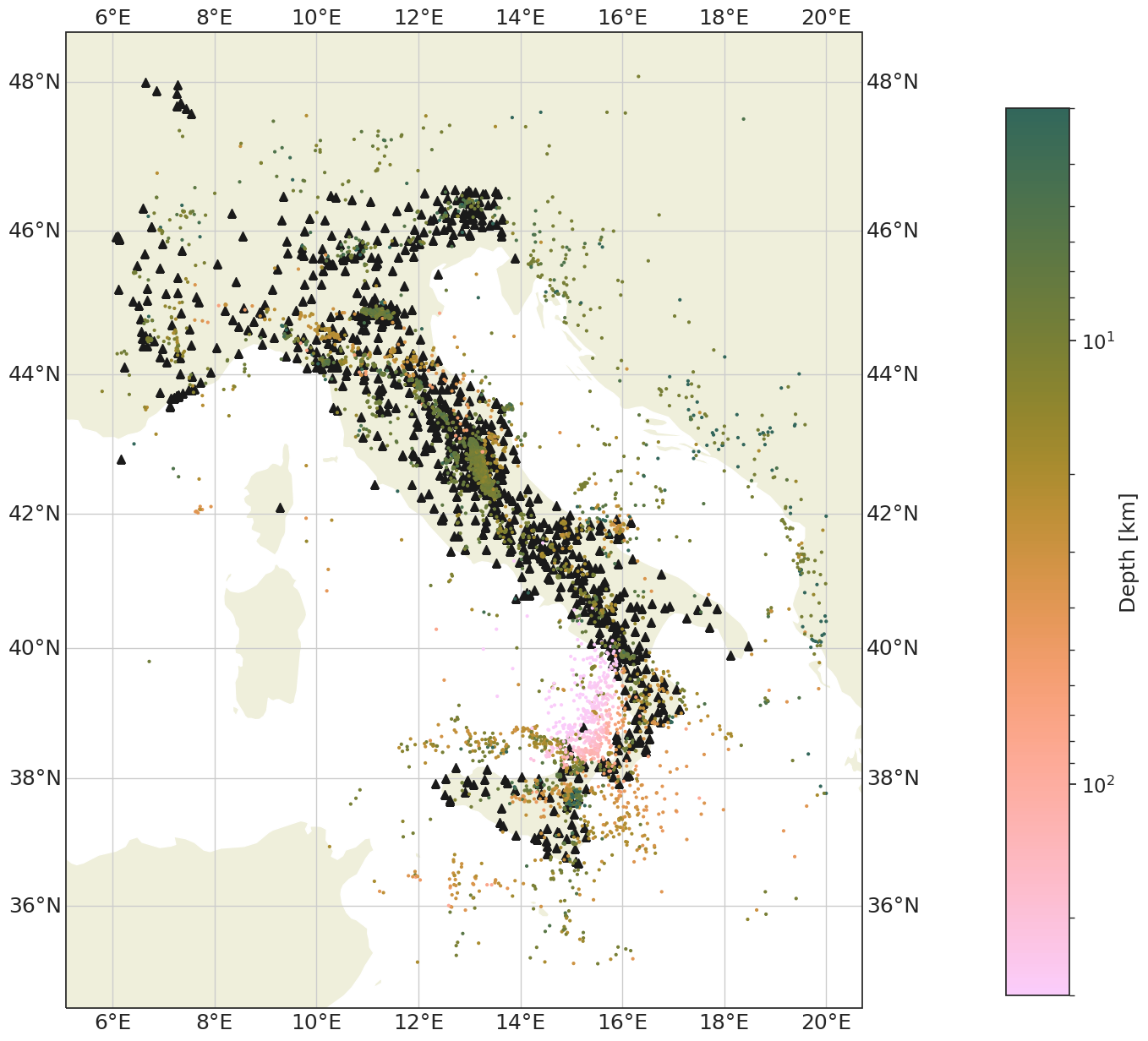}
 \caption{Event and station distribution for Italy. In the map, events are indicated by dots, stations by triangles. The event depth is encoded using color.}
 \label{SM-fig:highres_italy}
\end{figure}

\begin{figure}
 \includegraphics[width=\textwidth]{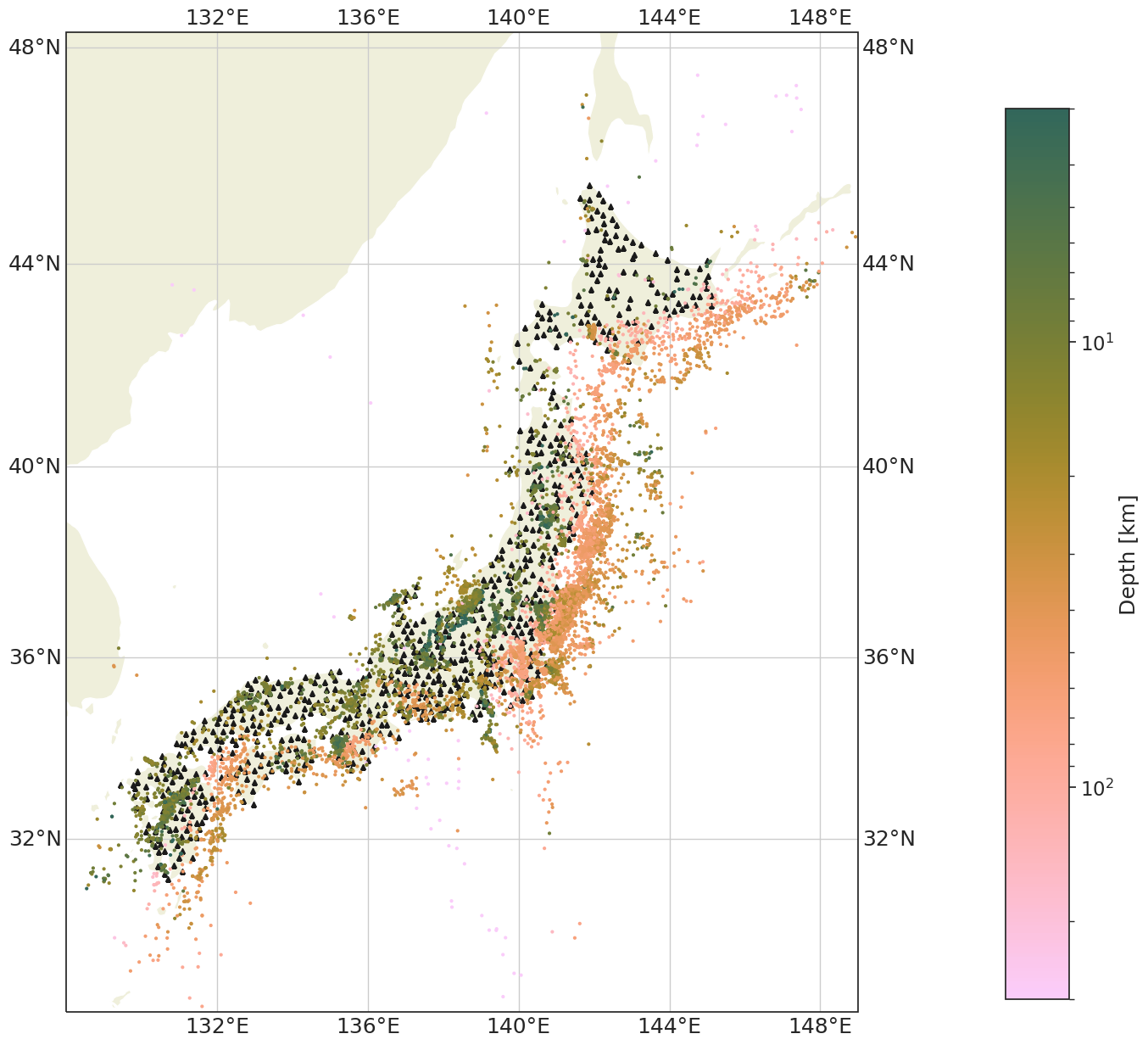}
 \caption{Event and station distribution for Japan. In the map, events are indicated by dots, stations by triangles. The event depth is encoded using color. There are $\sim$20 additional events far offshore in the catalog, which are outside the displayed map region.}
 \label{SM-fig:highres_japan}
\end{figure}

\begin{figure}
 \centering
 \includegraphics[width=\colwidth]{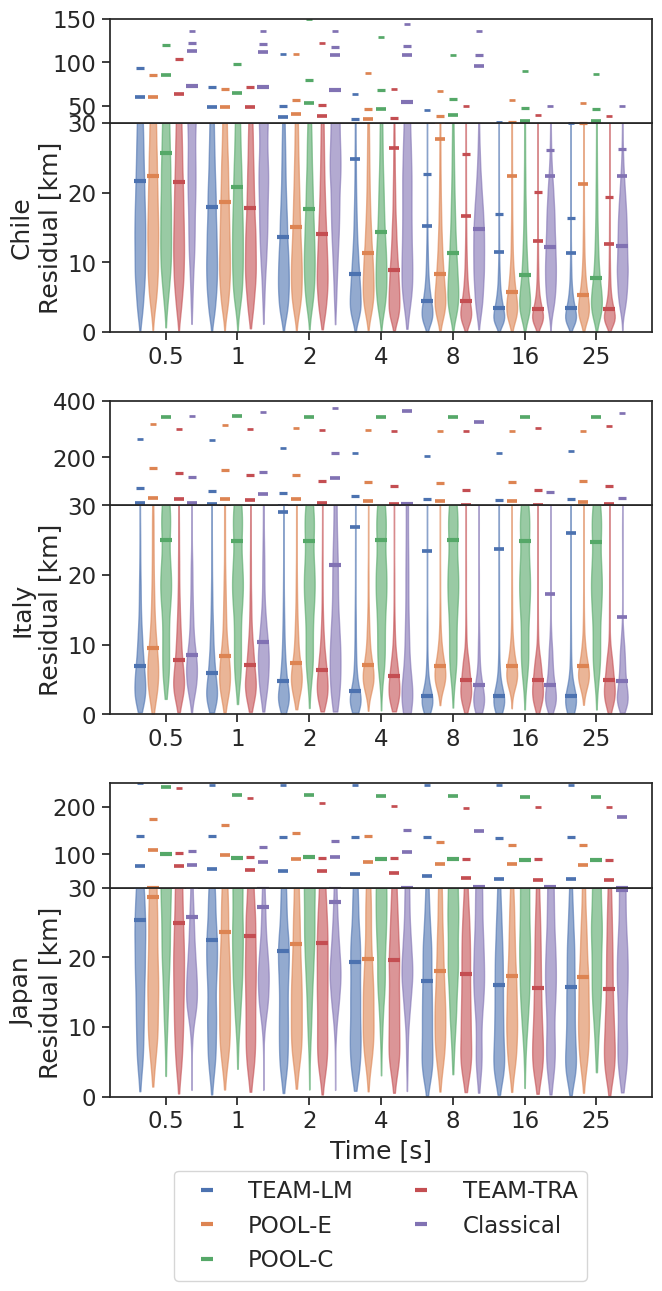}
 \caption{Distribution of the hypocentral errors for TEAM-LM, the pooling baseline with position embeddings (POOL-E), the pooling baseline with concatenated position (POOL-C), TEAM-LM with transfer learning (TEAM-TRA) and a classical baseline. Vertical lines mark the 50$^{th}$, 90$^{th}$, 95$^{th}$ and 99$^{th}$ error percentiles. The time indicates the time since the first P arrival at any station. We use the mean predictions.}
 \label{SM-fig:loc_residuals_hypo}
\end{figure}

\begin{figure}
 \includegraphics[width=\textwidth]{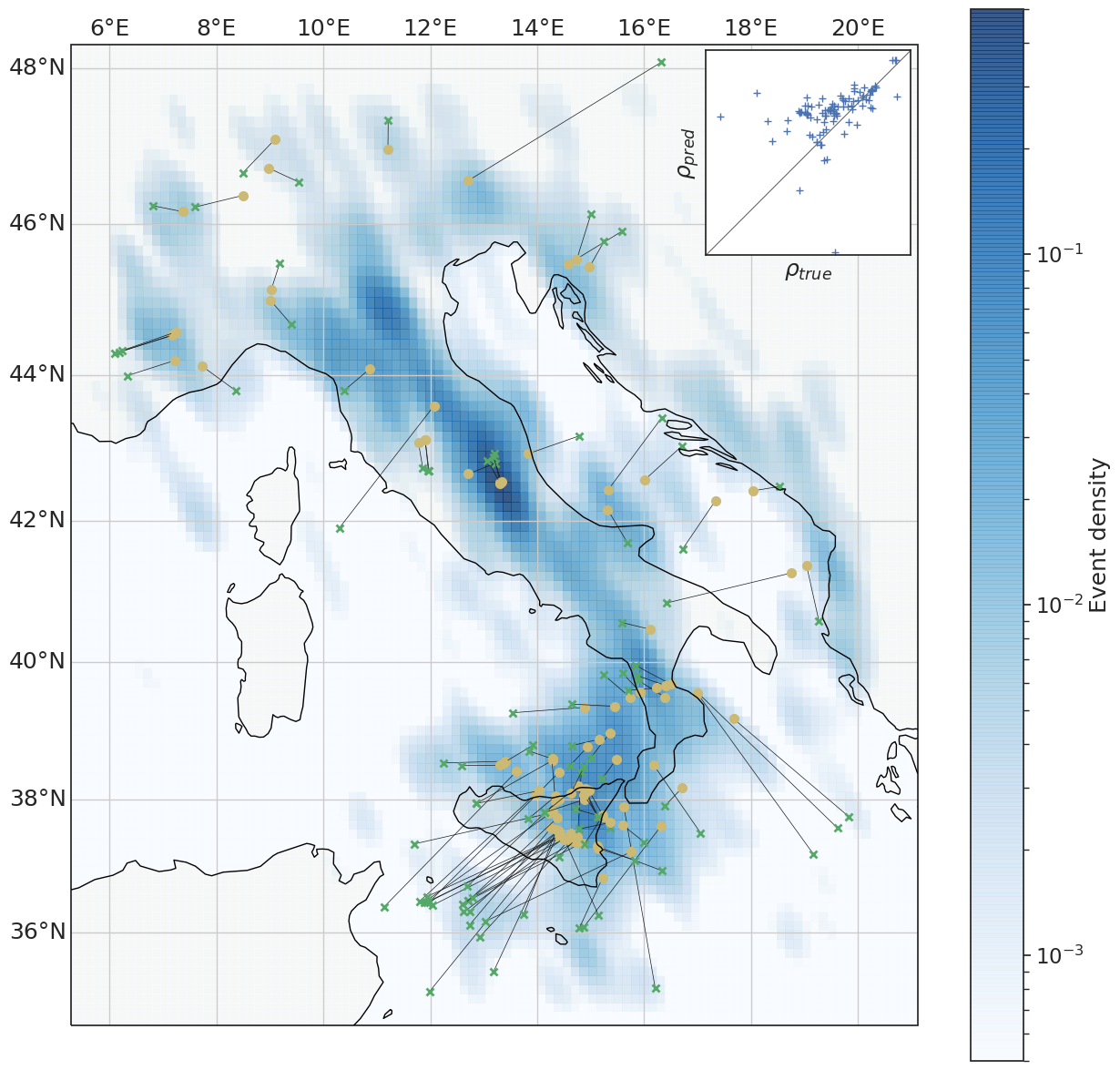}
 \caption{The 100 events with the highest location error in the Italy dataset overlayed on top of the spatial event density in the training dataset. The estimations use 16~s of data. Each event is denoted by a dot for the estimated location, a cross for the true location and a line connecting both. Stations are not shown as station coverage is dense. The event density is calculated using a Gaussian kernel density estimation and does not take into account the event depth. The inset shows the event density at the true event location in comparison to the event density at the predicted event location.}
 \label{SM-fig:mislocations_italy}
\end{figure}

\begin{figure}
 \includegraphics[width=\textwidth]{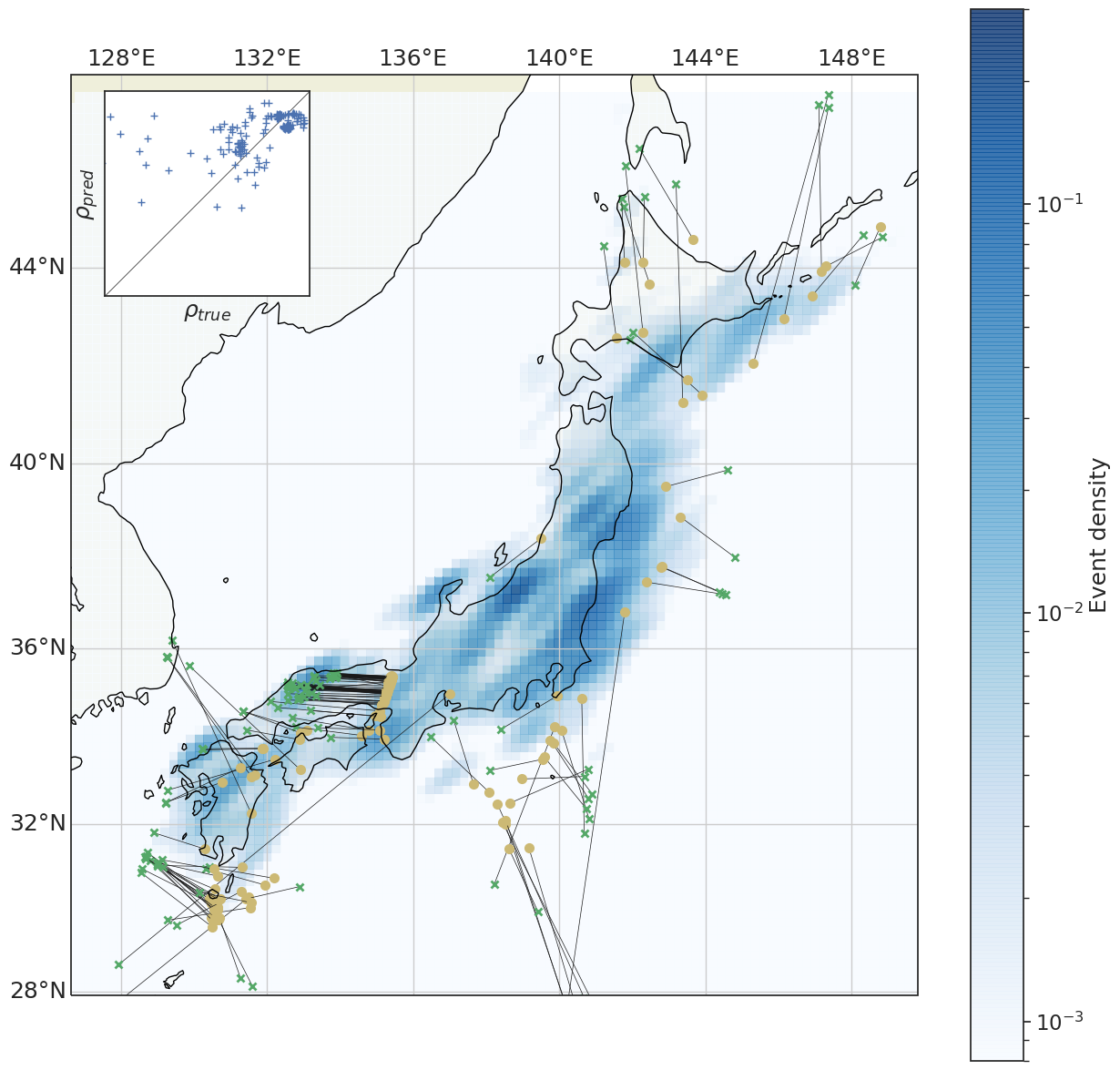}
 \caption{The 200 events with the highest location error in the Japan dataset overlayed on top of the spatial event density in the training dataset. The estimations use 16~s of data. Each event is denoted by a dot for the estimated location, a cross for the true location and a line connecting both. Stations are not shown as station coverage is dense. The event density is calculated using a Gaussian kernel density estimation and does not take into account the event depth. The inset shows the event density at the true event location in comparison to the event density at the predicted event location.}
 \label{SM-fig:mislocations_japan}
\end{figure}

\begin{figure}
 \centering
 \includegraphics[width=\textwidth]{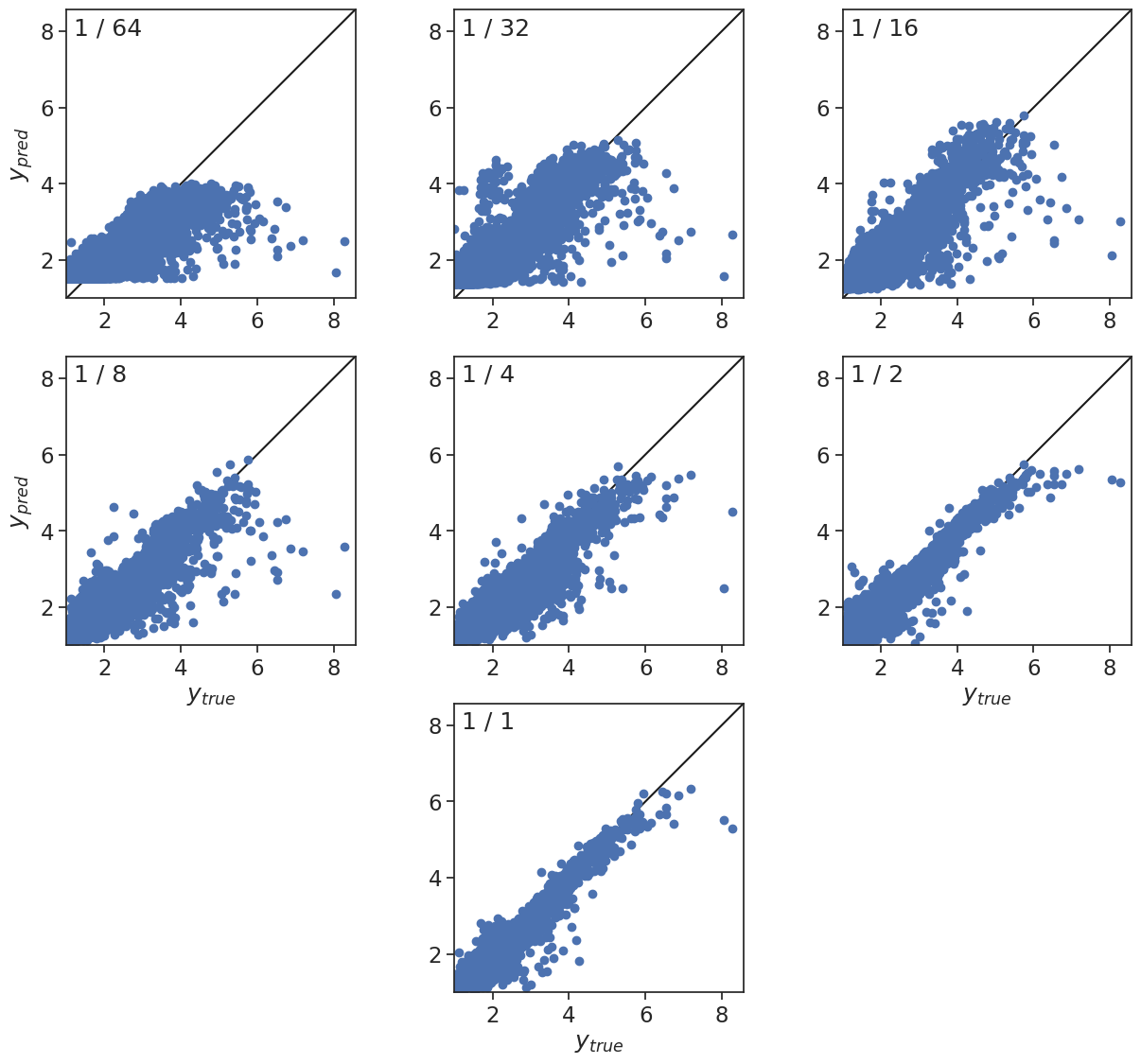}
 \caption{True and predicted magnitudes after 8 seconds using only parts of the datasets for training. All plots show the Chile dataset. The fraction in the corner indicates the amount of training and validation data used for model training. All models were evaluated on the full test dataset.}
 \label{SM-fig:data_sparsity_true_pred}
\end{figure}

\begin{table}
 \centering
 \caption{Experiment names for the results tables}
 \input{experiment_names.tex}
 \label{SM-tab:experiment_names}
\end{table}

\newgeometry{bottom=1cm}
\begin{landscape}
\thispagestyle{empty}
\begin{table}
 \centering
 \caption{Test set RMSE magnitude estimate across all magnitudes. For some experiments we additionally provide standard deviation. The standard deviations were obtained from six runs with different random model initialization. In this case the provided mean value is the mean over six runs. Note that the provided standard deviation denotes the empirical standard deviation of a single run, therefore the uncertainty of the mean expected to be smaller by a factor of $\sqrt{6}$. Due to computational constraints we are only able to provide standard deviations for a selected set of experiments.}
 \input{stats_rmse_test.tex}
 \label{SM-tab:rmse_stats}
\end{table}
\end{landscape}

\begin{landscape}
\thispagestyle{empty}
\begin{table}
 \centering
 \caption{Test set mean absolute error (MAE) magnitude estimate across all magnitudes}
 \input{stats_mae_test.tex}
 \label{SM-tab:mae_stats}
\end{table}
\end{landscape}

\begin{landscape}
\thispagestyle{empty}
\begin{table}
 \centering
 \caption{Test set R2 score across all magnitudes}
 \input{stats_r2_score_test.tex}
 \label{SM-tab:r2_stats}
\end{table}
\end{landscape}


\begin{landscape}
\thispagestyle{empty}
\begin{table}
 \centering
 \caption{Test set test statistic $d_\alpha$ for the Kolmogorov-Smirnov test across all magnitudes.}
 \input{stats_kolmogorow_smirnow_test.tex}
 \label{SM-tab:ks_stats}
\end{table}
\end{landscape}

\begin{landscape}
\thispagestyle{empty}
\begin{table}
 \centering
 \caption{Test set RMSE of magnitude estimate for large events}
 \input{stats_rmse_cut_test.tex}
 \label{SM-tab:rmse_stats_cut}
\end{table}
\end{landscape}

\begin{landscape}
\thispagestyle{empty}
\begin{table}
 \centering
 \caption{Test set MAE of magnitude estimate for large events}
 \input{stats_mae_cut_test.tex}
 \label{SM-tab:mae_stats_cut}
\end{table}
\end{landscape}

\begin{landscape}
\thispagestyle{empty}
\begin{table}
 \centering
 \caption{Test set R2 score of magnitude estimate for large events}
 \input{stats_r2_score_cut_test.tex}
 \label{SM-tab:r2_stats_cut}
\end{table}
\end{landscape}


\begin{landscape}
\thispagestyle{empty}
\begin{table}
 \centering
 \caption{Test set root squared mean for hypocentral error. We note that only 4 out of 10 models for the Italy location ensemble converged. We used only the converged models for the ensemble evaluation.}
 \input{loc_stats_rmse_hypo_test.tex}
 \label{SM-tab:rmse_stats_hypo}
\end{table}
\end{landscape}

\begin{landscape}
\thispagestyle{empty}
\begin{table}
 \centering
 \caption{Test set mean absolute hypocentral error. We note that only 4 out of 10 models for the Italy location ensemble converged. We used only the converged models for the ensemble evaluation.}
 \input{loc_stats_mae_hypo_test.tex}
 \label{SM-tab:mae_stats_hypo}
\end{table}
\end{landscape}

\begin{landscape}
\thispagestyle{empty}
\begin{table}
 \centering
 \caption{Test set root squared mean for epicentral error. We note that only 4 out of 10 models for the Italy location ensemble converged. We used only the converged models for the ensemble evaluation.}
 \input{loc_stats_rmse_epi_test.tex}
 \label{SM-tab:rmse_stats_epi}
\end{table}
\end{landscape}

\begin{landscape}
\thispagestyle{empty}
\begin{table}
 \centering
 \caption{Test set mean absolute epicentral error. We note that only 4 out of 10 models for the Italy location ensemble converged. We used only the converged models for the ensemble evaluation.}
 \input{loc_stats_mae_epi_test.tex}
 \label{SM-tab:mae_stats_epi}
\end{table}
\end{landscape}

\restoregeometry

\end{document}

%% file: datasets.tex
\begin{tabular}{l|c|c|c}
& Chile & Italy & Japan \\
\hline
Years & 2007 - 2014 & 2008 - 2019 & 1997 - 2018 \\
Training & 01/2007-08/2011 & 01/2008 - 12/2015 & 01/1997 - 03/2012 \\
& & \& 01/2017 - 12/2019 & \\
Test & 08/2012 - 12/2014 & 01/2016 - 12/2016 & 08/2013 - 12/2018 \\
Magnitudes &  1.21 - 8.27 & 2.7 - 6.5 & 2.7 - 9.0 \\
Magnitude scale & $M_A$ & $M_L$, $M_W$, $m_b$ & $M_\mathrm{JMA}$ \\
Depth [km] & 0 - 102 - 183 & 0 - 10 - 617& 0 - 19 - 682\\
Distance [km] & 0.1 - 180 - 640 & 0.1 - 180 - 630 & 0.2 - 120 - 3190 \\
Events & 96,133 & 7,055 & 13,512 \\
Unique stations & 24 & 1,080 & 697 \\
Traces & 1,605,983 & 494,183 & 372,661 \\
Traces per event & 16.7 & 70.3 & 27.6 \\
Sensor type & BB & SM & SM \& SM-borehole \\
Catalog source & \cite{munchmeyerLowUncertaintyMultifeature2020} & INGV & NIED \\
\end{tabular}

%% file: seismic_networks.tex
\begin{tabular}{p{2cm}p{1.5cm}p{13cm}}
 Region & Network & Reference \\
 \hline
 Chile
 & GE & \cite{fdsnGE} \\
 & C, C1 & \cite{fdsnC1} \\
 & 8F & \cite{fdsn8F} \\
 & IQ & \cite{fdsnIQ} \\
 & 5E & \cite{fdsn5E} \\
 Italy
 & 3A & \cite{network_3a} \\
 & BA & \cite{network_ba} \\
 & FR & \cite{network_fr} \\
 & GU & \cite{network_gu} \\
 & IT & \cite{network_it} \\
 & IV & \cite{network_iv} \\
 & IX & \cite{network_ix} \\
 & MN & \cite{network_mn} \\
 & NI & \cite{network_ni} \\
 & OX & \cite{network_ox} \\
 & RA & \cite{network_ra} \\
 & ST & \cite{network_st} \\
 & TV & \cite{network_tv} \\
 & XO & \cite{network_xo} \\
 Japan
 & KiK-Net & \cite{nied2019kiknet}
\end{tabular}

%% file: convolutions.tex
\begin{tabular}{lccc}
Layer & Filters & Kernel size & Stride \\
\hline
Conv2D & 8 & 5, 1 & 5, 1 \\
Conv2D & 32 & 16, 3 & 1, 3 \\
Flatten to 1D & & & \\
Conv1D & 64 & 16 & 1 \\
MaxPool1D & & 2 & 2 \\
Conv1D & 128 & 16 & 1 \\
MaxPool1D & & 2 & 2 \\
Conv1D & 32 & 8 & 1 \\
MaxPool1D & & 2 & 2 \\
Conv1D & 32 & 8 & 1 \\
Conv1D & 16 & 4 & 1 \\
Flatten to 0D& & & \\
Concatenate scale & & & \\
FC & 500 & & \\
FC & 500 & & \\
FC & 500 & & \\
\end{tabular}

%% file: transformer.tex
\begin{tabular}{lc}
Feature & Value \\
\hline
\# Layers & 6 \\
Dimension & 500 \\
Feed forward dimension & 1000 \\
\# Heads & 10 \\
Maximum number of stations & 25 \\
Dropout & 0 \\
Activation & GeLu \\
\end{tabular}

%% file: experiment_names.tex
\begin{tabular}{ll}
Name & Explanation \\
\hline
Baseline & Baseline method \\
Plain & Model trained with a single loss for magnitude or location \\
Multi-task & Model trained with both a loss for magnitude and for location \\
High magnitudes & Model only trained on events with magnitude above 4.5 (Chile), 3.5 (Italy), 5.5 (Japan) \\
Transfer & Model trained with transfer learning and a single loss \\
Multi-task transfer & Model trained with transfer learning and both losses \\
Joint & Model trained on all datasets jointly with a single loss\\
Joint multi-task & Model trained on all datasets jointly with both losses \\
Velocity & Model trained with acceleration traces integrated to velocity \\
Ensemble & Model trained as an ensemble of size 10 \\
No upsampling & Model trained without upsampling large magnitudes \\
Pooling (Emb) & Model trained with the pooling architecture using position embeddings \\
Pooling (Concat) & Model trained with the pooling architecture and concatenated coordinates \\
\end{tabular}

%% file: stats_rmse_test.tex
\resizebox{1.7\textwidth}{!}{%
\begin{tabular}{c|ccccccc|ccccccc|ccccccc}
 & \multicolumn{7}{|c}{Chile} & \multicolumn{7}{|c}{Italy} & \multicolumn{7}{|c}{Japan} \\
 & 0.5s & 1.0s & 2.0s & 4.0s & 8.0s & 16.0s & 25.0s & 0.5s & 1.0s & 2.0s & 4.0s & 8.0s & 16.0s & 25.0s & 0.5s & 1.0s & 2.0s & 4.0s & 8.0s & 16.0s & 25.0s \\
\hline
Baseline & - & 0.55 & 0.52 & 0.47 & 0.42 & 0.42 & 0.43 & - & 0.77 & 0.68 & 0.46 & 0.35 & 0.35 & 0.35 & - & 0.65 & 0.51 & 0.44 & 0.38 & 0.31 & 0.30 \\
Multi-task & 0.31 & 0.25 & \textbf{0.20} & 0.15 & \textbf{0.11} & \textbf{0.08} & \textbf{0.07} & 0.36 & 0.34 & 0.30 & 0.26 & 0.24 & 0.23 & 0.23 & 0.65 & 0.47 & 0.38 & 0.33 & 0.27 & 0.23 & 0.23 \\
High magnitudes & 0.72 & 0.72 & 0.72 & 0.72 & 0.72 & 0.72 & 0.72 & 0.46 & 0.46 & 0.45 & 0.45 & 0.44 & 0.43 & 0.43 & \textbf{0.51} & 0.50 & 0.47 & 0.43 & 0.38 & 0.34 & 0.36 \\
Plain & 0.31 & 0.25 & 0.21 & 0.16 & 0.12 & 0.09 & 0.08 & 0.34 & 0.33 & 0.29 & 0.24 & 0.21 & 0.20 & 0.20 & 0.65 & 0.47 & 0.37 & 0.33 & 0.26 & 0.22 & 0.22 \\
 & $\pm$0.02 & $\pm$0.02 & $\pm$0.02 & $\pm$0.01 & $\pm$0.00 & $\pm$0.00 & $\pm$0.00 & $\pm$0.01 & $\pm$0.01 & $\pm$0.01 & $\pm$0.02 & $\pm$0.02 & $\pm$0.02 & $\pm$0.02 & $\pm$0.02 & $\pm$0.01 & $\pm$0.01 & $\pm$0.01 & $\pm$0.01 & $\pm$0.01 & $\pm$0.01 \\
Ensemble & \textbf{0.29} & \textbf{0.23} & \textbf{0.19} & \textbf{0.15} & \textbf{0.11} & 0.08 & 0.08 & \textbf{0.32} & \textbf{0.31} & \textbf{0.26} & 0.21 & 0.18 & 0.17 & 0.18 & 0.64 & \textbf{0.43} & \textbf{0.35} & \textbf{0.30} & \textbf{0.24} & \textbf{0.21} & \textbf{0.21} \\
No Upsampling & 0.31 & 0.25 & 0.21 & 0.16 & 0.12 & 0.09 & 0.08 & \textbf{0.33} & \textbf{0.31} & \textbf{0.28} & 0.23 & 0.19 & 0.18 & 0.18 & 0.54 & \textbf{0.42} & \textbf{0.35} & \textbf{0.31} & \textbf{0.25} & \textbf{0.21} & \textbf{0.22} \\
Pooling (Emb) & 0.31 & 0.25 & 0.21 & 0.16 & 0.13 & 0.09 & 0.09 & 0.36 & 0.33 & 0.29 & 0.25 & 0.21 & 0.20 & 0.21 & 0.66 & 0.49 & 0.41 & 0.36 & 0.31 & 0.28 & 0.28 \\
 & $\pm$0.01 & $\pm$0.00 & $\pm$0.00 & $\pm$0.00 & $\pm$0.00 & $\pm$0.00 & $\pm$0.00 & $\pm$0.02 & $\pm$0.02 & $\pm$0.02 & $\pm$0.03 & $\pm$0.03 & $\pm$0.03 & $\pm$0.03 & $\pm$0.03 & $\pm$0.05 & $\pm$0.05 & $\pm$0.05 & $\pm$0.06 & $\pm$0.07 & $\pm$0.07 \\
Pooling (Concat) & 0.35 & 0.28 & 0.24 & 0.20 & 0.16 & 0.11 & 0.11 & 0.37 & 0.33 & 0.29 & 0.24 & 0.21 & 0.21 & 0.22 & 0.71 & 0.50 & 0.41 & 0.35 & 0.29 & 0.26 & 0.26 \\
 & $\pm$0.01 & $\pm$0.01 & $\pm$0.02 & $\pm$0.01 & $\pm$0.01 & $\pm$0.01 & $\pm$0.01 & $\pm$0.03 & $\pm$0.02 & $\pm$0.01 & $\pm$0.01 & $\pm$0.01 & $\pm$0.02 & $\pm$0.02 & $\pm$0.01 & $\pm$0.02 & $\pm$0.03 & $\pm$0.02 & $\pm$0.03 & $\pm$0.04 & $\pm$0.04 \\
Transfer & 0.32 & 0.25 & 0.20 & 0.15 & 0.12 & 0.09 & 0.08 & 0.36 & 0.33 & 0.28 & 0.22 & 0.19 & 0.19 & 0.22 & 0.71 & 0.52 & 0.40 & 0.35 & 0.30 & 0.25 & 0.25 \\
Multi-task transfer & 0.31 & \textbf{0.24} & \textbf{0.19} & \textbf{0.14} & \textbf{0.11} & \textbf{0.08} & \textbf{0.07} & 0.36 & 0.33 & \textbf{0.27} & \textbf{0.20} & \textbf{0.17} & \textbf{0.17} & \textbf{0.17} & 0.63 & 0.47 & 0.39 & 0.34 & 0.27 & 0.23 & 0.23 \\
Velocity & - & - & - & - & - & - & - & 0.37 & 0.35 & 0.28 & 0.22 & 0.18 & 0.18 & 0.19 & 0.64 & 0.47 & 0.38 & 0.35 & 0.29 & 0.24 & 0.25 \\
Joint multi-task & 0.32 & 0.25 & 0.20 & 0.15 & 0.11 & 0.08 & 0.08 & 0.35 & \textbf{0.32} & \textbf{0.27} & \textbf{0.21} & \textbf{0.17} & \textbf{0.16} & \textbf{0.17} & 0.65 & 0.50 & 0.40 & 0.35 & 0.28 & 0.24 & 0.25 \\
Joint & 0.34 & 0.27 & 0.22 & 0.18 & 0.14 & 0.11 & 0.10 & 0.34 & \textbf{0.32} & \textbf{0.27} & \textbf{0.21} & \textbf{0.17} & 0.17 & 0.18 & 0.62 & 0.49 & 0.40 & 0.35 & 0.30 & 0.25 & 0.25 \\
\end{tabular}}

%% file: stats_mae_test.tex
\resizebox{1.7\textwidth}{!}{%
\begin{tabular}{c|ccccccc|ccccccc|ccccccc}
 & \multicolumn{7}{|c}{Chile} & \multicolumn{7}{|c}{Italy} & \multicolumn{7}{|c}{Japan} \\
 & 0.5s & 1.0s & 2.0s & 4.0s & 8.0s & 16.0s & 25.0s & 0.5s & 1.0s & 2.0s & 4.0s & 8.0s & 16.0s & 25.0s & 0.5s & 1.0s & 2.0s & 4.0s & 8.0s & 16.0s & 25.0s \\
\hline
Baseline & - & 0.43 & 0.41 & 0.37 & 0.33 & 0.33 & 0.34 & - & 0.63 & 0.53 & 0.34 & 0.25 & 0.24 & 0.24 & - & 0.50 & 0.40 & 0.34 & 0.28 & 0.23 & 0.22 \\
Multi-task & 0.23 & 0.18 & \textbf{0.14} & \textbf{0.10} & \textbf{0.07} & \textbf{0.04} & \textbf{0.04} & 0.26 & 0.24 & 0.21 & 0.18 & 0.16 & 0.15 & 0.15 & 0.48 & 0.33 & 0.26 & 0.21 & 0.17 & 0.15 & 0.15 \\
High magnitudes & 0.53 & 0.52 & 0.52 & 0.52 & 0.52 & 0.52 & 0.52 & 0.30 & 0.30 & 0.29 & 0.27 & 0.26 & 0.26 & 0.26 & \textbf{0.40} & 0.39 & 0.37 & 0.32 & 0.30 & 0.27 & 0.28 \\
Plain & \textbf{0.23} & 0.18 & 0.14 & 0.11 & 0.08 & 0.05 & 0.04 & 0.26 & 0.24 & 0.21 & 0.17 & 0.15 & 0.15 & 0.15 & 0.50 & 0.34 & 0.26 & 0.22 & 0.17 & 0.15 & 0.14 \\
 & $\pm$0.01 & $\pm$0.01 & $\pm$0.01 & $\pm$0.01 & $\pm$0.00 & $\pm$0.00 & $\pm$0.00 & $\pm$0.01 & $\pm$0.01 & $\pm$0.02 & $\pm$0.02 & $\pm$0.02 & $\pm$0.02 & $\pm$0.02 & $\pm$0.01 & $\pm$0.01 & $\pm$0.01 & $\pm$0.01 & $\pm$0.01 & $\pm$0.01 & $\pm$0.01 \\
Ensemble & \textbf{0.22} & \textbf{0.17} & \textbf{0.13} & \textbf{0.10} & 0.07 & 0.05 & 0.04 & \textbf{0.24} & 0.23 & 0.19 & 0.15 & 0.13 & 0.12 & 0.13 & 0.48 & \textbf{0.31} & \textbf{0.24} & \textbf{0.19} & \textbf{0.15} & \textbf{0.13} & \textbf{0.13} \\
No Upsampling & \textbf{0.23} & 0.18 & 0.14 & 0.11 & 0.08 & 0.05 & 0.04 & \textbf{0.24} & 0.23 & 0.20 & 0.16 & 0.14 & 0.12 & 0.12 & \textbf{0.40} & \textbf{0.30} & \textbf{0.24} & 0.20 & 0.16 & 0.14 & 0.14 \\
Pooling (Emb) & 0.23 & 0.18 & 0.15 & 0.11 & 0.08 & 0.06 & 0.05 & 0.27 & 0.25 & 0.22 & 0.18 & 0.15 & 0.15 & 0.15 & 0.51 & 0.37 & 0.31 & 0.26 & 0.23 & 0.21 & 0.20 \\
 & $\pm$0.01 & $\pm$0.01 & $\pm$0.01 & $\pm$0.01 & $\pm$0.01 & $\pm$0.01 & $\pm$0.01 & $\pm$0.02 & $\pm$0.02 & $\pm$0.02 & $\pm$0.02 & $\pm$0.03 & $\pm$0.02 & $\pm$0.02 & $\pm$0.04 & $\pm$0.05 & $\pm$0.06 & $\pm$0.06 & $\pm$0.07 & $\pm$0.08 & $\pm$0.08 \\
Pooling (Concat) & 0.25 & 0.20 & 0.16 & 0.13 & 0.10 & 0.07 & 0.06 & 0.28 & 0.25 & 0.21 & 0.17 & 0.16 & 0.15 & 0.16 & 0.55 & 0.38 & 0.29 & 0.25 & 0.20 & 0.18 & 0.18 \\
 & $\pm$0.01 & $\pm$0.01 & $\pm$0.01 & $\pm$0.01 & $\pm$0.01 & $\pm$0.01 & $\pm$0.01 & $\pm$0.02 & $\pm$0.01 & $\pm$0.01 & $\pm$0.01 & $\pm$0.02 & $\pm$0.02 & $\pm$0.02 & $\pm$0.01 & $\pm$0.02 & $\pm$0.03 & $\pm$0.03 & $\pm$0.04 & $\pm$0.05 & $\pm$0.05 \\
Transfer & 0.24 & 0.18 & 0.14 & 0.11 & 0.08 & 0.05 & 0.05 & 0.26 & 0.24 & 0.20 & 0.16 & 0.14 & 0.14 & 0.15 & 0.53 & 0.37 & 0.27 & 0.22 & 0.18 & 0.15 & 0.15 \\
Multi-task transfer & 0.23 & \textbf{0.18} & \textbf{0.13} & \textbf{0.10} & \textbf{0.07} & \textbf{0.04} & \textbf{0.04} & \textbf{0.24} & \textbf{0.22} & \textbf{0.18} & \textbf{0.13} & \textbf{0.11} & \textbf{0.11} & \textbf{0.11} & 0.46 & 0.33 & 0.26 & 0.21 & 0.17 & 0.14 & 0.14 \\
Velocity & - & - & - & - & - & - & - & 0.25 & 0.24 & 0.19 & 0.14 & 0.12 & 0.13 & 0.13 & 0.47 & 0.33 & 0.26 & 0.21 & 0.17 & 0.14 & 0.14 \\
Joint multi-task & 0.23 & 0.18 & 0.14 & 0.10 & 0.07 & 0.05 & 0.04 & \textbf{0.24} & \textbf{0.21} & \textbf{0.18} & \textbf{0.13} & \textbf{0.11} & \textbf{0.11} & \textbf{0.11} & 0.48 & 0.35 & 0.27 & 0.22 & 0.17 & 0.15 & 0.15 \\
Joint & 0.26 & 0.20 & 0.16 & 0.13 & 0.10 & 0.08 & 0.07 & \textbf{0.25} & 0.23 & 0.19 & 0.14 & \textbf{0.12} & 0.12 & 0.12 & 0.46 & 0.35 & 0.28 & 0.22 & 0.18 & 0.16 & 0.16 \\
\end{tabular}}

%% file: stats_r2_score_test.tex
\resizebox{1.7\textwidth}{!}{%
\begin{tabular}{c|ccccccc|ccccccc|ccccccc}
 & \multicolumn{7}{|c}{Chile} & \multicolumn{7}{|c}{Italy} & \multicolumn{7}{|c}{Japan} \\
 & 0.5s & 1.0s & 2.0s & 4.0s & 8.0s & 16.0s & 25.0s & 0.5s & 1.0s & 2.0s & 4.0s & 8.0s & 16.0s & 25.0s & 0.5s & 1.0s & 2.0s & 4.0s & 8.0s & 16.0s & 25.0s \\
\hline
Baseline & - & 0.53 & 0.56 & 0.60 & 0.62 & 0.59 & 0.57 & - & -2.26 & -1.83 & -0.34 & 0.21 & 0.14 & 0.10 & - & 0.26 & 0.54 & 0.66 & 0.75 & 0.82 & 0.84 \\
Multi-task & 0.74 & 0.84 & \textbf{0.90} & \textbf{0.94} & \textbf{0.97} & \textbf{0.98} & \textbf{0.99} & 0.17 & 0.25 & 0.41 & 0.55 & 0.64 & 0.65 & 0.66 & 0.31 & 0.64 & 0.76 & 0.82 & 0.88 & \textbf{0.91} & \textbf{0.91} \\
High magnitudes & 0.03 & 0.03 & 0.02 & 0.03 & 0.03 & 0.02 & 0.01 & -0.29 & -0.29 & -0.27 & -0.26 & -0.17 & -0.14 & -0.15 & 0.02 & 0.07 & 0.17 & 0.32 & 0.45 & 0.57 & 0.50 \\
Plain & 0.74 & 0.83 & 0.89 & \textbf{0.93} & \textbf{0.96} & \textbf{0.98} & \textbf{0.98} & 0.24 & 0.29 & 0.45 & 0.62 & 0.72 & 0.74 & 0.73 & 0.29 & 0.64 & 0.77 & 0.82 & \textbf{0.89} & \textbf{0.92} & \textbf{0.92} \\
 & $\pm$0.03 & $\pm$0.02 & $\pm$0.02 & $\pm$0.01 & $\pm$0.00 & $\pm$0.00 & $\pm$0.00 & $\pm$0.05 & $\pm$0.04 & $\pm$0.04 & $\pm$0.05 & $\pm$0.05 & $\pm$0.04 & $\pm$0.04 & $\pm$0.04 & $\pm$0.01 & $\pm$0.01 & $\pm$0.01 & $\pm$0.01 & $\pm$0.00 & $\pm$0.00 \\
Ensemble & \textbf{0.77} & \textbf{0.86} & \textbf{0.91} & \textbf{0.94} & \textbf{0.97} & \textbf{0.98} & \textbf{0.98} & \textbf{0.33} & \textbf{0.39} & \textbf{0.55} & 0.70 & 0.79 & 0.81 & 0.80 & 0.33 & \textbf{0.70} & \textbf{0.80} & \textbf{0.85} & \textbf{0.90} & \textbf{0.93} & \textbf{0.93} \\
No Upsampling & 0.74 & 0.83 & 0.88 & \textbf{0.93} & \textbf{0.96} & \textbf{0.98} & \textbf{0.98} & 0.29 & 0.37 & 0.50 & 0.65 & 0.76 & 0.79 & 0.79 & \textbf{0.52} & \textbf{0.71} & \textbf{0.79} & \textbf{0.84} & \textbf{0.89} & \textbf{0.92} & \textbf{0.92} \\
Pooling (Emb) & 0.74 & 0.83 & 0.88 & \textbf{0.93} & \textbf{0.96} & \textbf{0.98} & \textbf{0.98} & 0.17 & 0.28 & 0.44 & 0.61 & 0.71 & 0.73 & 0.71 & 0.27 & 0.60 & 0.72 & 0.78 & 0.84 & 0.86 & 0.86 \\
 & $\pm$0.01 & $\pm$0.00 & $\pm$0.00 & $\pm$0.00 & $\pm$0.00 & $\pm$0.00 & $\pm$0.00 & $\pm$0.11 & $\pm$0.09 & $\pm$0.09 & $\pm$0.09 & $\pm$0.10 & $\pm$0.08 & $\pm$0.08 & $\pm$0.07 & $\pm$0.08 & $\pm$0.07 & $\pm$0.06 & $\pm$0.06 & $\pm$0.07 & $\pm$0.07 \\
Pooling (Concat) & 0.68 & 0.78 & 0.85 & 0.90 & 0.94 & \textbf{0.96} & \textbf{0.97} & 0.13 & 0.28 & 0.46 & 0.62 & 0.70 & 0.72 & 0.69 & 0.17 & 0.58 & 0.73 & 0.79 & 0.86 & 0.88 & 0.89 \\
 & $\pm$0.02 & $\pm$0.02 & $\pm$0.02 & $\pm$0.01 & $\pm$0.01 & $\pm$0.01 & $\pm$0.01 & $\pm$0.14 & $\pm$0.07 & $\pm$0.03 & $\pm$0.03 & $\pm$0.04 & $\pm$0.06 & $\pm$0.06 & $\pm$0.03 & $\pm$0.03 & $\pm$0.04 & $\pm$0.03 & $\pm$0.03 & $\pm$0.04 & $\pm$0.04 \\
Transfer & 0.73 & 0.83 & \textbf{0.89} & \textbf{0.94} & \textbf{0.96} & \textbf{0.98} & \textbf{0.98} & 0.18 & 0.28 & 0.49 & 0.67 & 0.76 & 0.76 & 0.69 & 0.18 & 0.56 & 0.73 & 0.80 & 0.86 & 0.90 & 0.90 \\
Multi-task transfer & 0.75 & \textbf{0.84} & \textbf{0.91} & \textbf{0.95} & \textbf{0.97} & \textbf{0.98} & \textbf{0.99} & 0.16 & 0.30 & 0.53 & \textbf{0.73} & \textbf{0.81} & \textbf{0.82} & \textbf{0.81} & 0.35 & 0.64 & 0.75 & 0.81 & 0.88 & \textbf{0.91} & \textbf{0.91} \\
Velocity & - & - & - & - & - & - & - & 0.11 & 0.21 & 0.48 & 0.69 & 0.79 & 0.78 & 0.77 & 0.32 & 0.63 & 0.76 & 0.80 & 0.86 & 0.90 & 0.90 \\
Joint multi-task & 0.73 & 0.83 & \textbf{0.89} & \textbf{0.94} & \textbf{0.97} & \textbf{0.98} & \textbf{0.98} & 0.22 & 0.35 & \textbf{0.54} & \textbf{0.72} & \textbf{0.81} & \textbf{0.83} & \textbf{0.82} & 0.31 & 0.59 & 0.73 & 0.80 & 0.87 & 0.90 & 0.90 \\
Joint & 0.69 & 0.80 & 0.87 & 0.92 & 0.95 & \textbf{0.97} & \textbf{0.97} & 0.25 & 0.34 & 0.51 & 0.72 & \textbf{0.82} & 0.81 & 0.78 & 0.37 & 0.60 & 0.74 & 0.80 & 0.85 & 0.90 & 0.89 \\
\end{tabular}}

%% file: stats_kolmogorow_smirnow_test.tex
\resizebox{1.7\textwidth}{!}{%
\begin{tabular}{c|ccccccc|ccccccc|ccccccc}
 & \multicolumn{7}{|c}{Chile} & \multicolumn{7}{|c}{Italy} & \multicolumn{7}{|c}{Japan} \\
 & 0.5s & 1.0s & 2.0s & 4.0s & 8.0s & 16.0s & 25.0s & 0.5s & 1.0s & 2.0s & 4.0s & 8.0s & 16.0s & 25.0s & 0.5s & 1.0s & 2.0s & 4.0s & 8.0s & 16.0s & 25.0s \\
\hline
Baseline & - & - & - & - & - & - & - & - & - & - & - & - & - & - & - & - & - & - & - & - & - \\
Multi-task & 0.05 & 0.03 & 0.03 & 0.06 & 0.04 & 0.03 & 0.04 & 0.09 & 0.10 & 0.14 & 0.16 & 0.17 & 0.22 & 0.22 & 0.22 & 0.17 & 0.17 & 0.18 & 0.19 & 0.19 & 0.20 \\
High magnitudes & 0.15 & 0.16 & 0.15 & 0.15 & 0.14 & 0.13 & 0.14 & 0.13 & 0.16 & 0.19 & 0.22 & 0.22 & 0.25 & 0.24 & 0.16 & 0.14 & 0.12 & 0.09 & 0.12 & 0.11 & 0.11 \\
Plain & 0.04 & 0.04 & 0.04 & 0.05 & 0.04 & 0.04 & 0.03 & 0.11 & 0.13 & 0.16 & 0.15 & 0.16 & 0.15 & 0.16 & 0.19 & 0.13 & 0.12 & 0.13 & 0.14 & 0.12 & 0.11 \\
 & $\pm$0.02 & $\pm$0.02 & $\pm$0.02 & $\pm$0.02 & $\pm$0.01 & $\pm$0.01 & $\pm$0.01 & $\pm$0.03 & $\pm$0.04 & $\pm$0.05 & $\pm$0.07 & $\pm$0.06 & $\pm$0.08 & $\pm$0.07 & $\pm$0.04 & $\pm$0.04 & $\pm$0.05 & $\pm$0.06 & $\pm$0.07 & $\pm$0.06 & $\pm$0.05 \\
Ensemble & 0.04 & 0.04 & 0.03 & 0.03 & 0.03 & 0.04 & 0.04 & \textbf{0.06} & 0.07 & \textbf{0.06} & \textbf{0.07} & \textbf{0.09} & 0.09 & 0.10 & 0.19 & 0.12 & \textbf{0.09} & 0.08 & 0.09 & 0.08 & 0.08 \\
No Upsampling & 0.04 & 0.03 & 0.04 & 0.05 & 0.05 & 0.05 & 0.04 & 0.07 & 0.09 & 0.11 & 0.15 & 0.17 & \textbf{0.09} & \textbf{0.09} & \textbf{0.11} & \textbf{0.10} & \textbf{0.09} & 0.11 & 0.14 & 0.15 & 0.14 \\
Pooling (Emb) & 0.03 & 0.03 & 0.03 & 0.03 & 0.04 & 0.03 & 0.05 & 0.14 & 0.14 & 0.16 & 0.17 & 0.15 & 0.16 & 0.18 & 0.19 & 0.13 & 0.16 & 0.16 & 0.18 & 0.17 & 0.17 \\
 & $\pm$0.01 & $\pm$0.01 & $\pm$0.01 & $\pm$0.01 & $\pm$0.01 & $\pm$0.02 & $\pm$0.02 & $\pm$0.04 & $\pm$0.03 & $\pm$0.05 & $\pm$0.08 & $\pm$0.05 & $\pm$0.07 & $\pm$0.08 & $\pm$0.03 & $\pm$0.04 & $\pm$0.02 & $\pm$0.02 & $\pm$0.03 & $\pm$0.04 & $\pm$0.04 \\
Pooling (Concat) & 0.04 & 0.04 & 0.05 & 0.05 & 0.05 & 0.04 & 0.05 & 0.14 & 0.12 & 0.12 & 0.14 & 0.17 & 0.17 & 0.19 & 0.20 & 0.12 & 0.11 & 0.12 & 0.13 & 0.12 & 0.13 \\
 & $\pm$0.01 & $\pm$0.01 & $\pm$0.01 & $\pm$0.01 & $\pm$0.01 & $\pm$0.02 & $\pm$0.02 & $\pm$0.02 & $\pm$0.02 & $\pm$0.04 & $\pm$0.04 & $\pm$0.07 & $\pm$0.06 & $\pm$0.05 & $\pm$0.02 & $\pm$0.03 & $\pm$0.03 & $\pm$0.02 & $\pm$0.05 & $\pm$0.07 & $\pm$0.06 \\
Transfer & 0.05 & \textbf{0.02} & 0.03 & \textbf{0.02} & \textbf{0.02} & \textbf{0.02} & \textbf{0.02} & 0.08 & \textbf{0.06} & 0.10 & 0.15 & 0.25 & 0.24 & 0.30 & 0.17 & 0.15 & 0.15 & 0.15 & 0.14 & 0.11 & 0.10 \\
Multi-task transfer & \textbf{0.02} & 0.03 & \textbf{0.02} & 0.04 & 0.04 & 0.06 & 0.03 & 0.14 & 0.14 & 0.13 & 0.12 & 0.18 & 0.20 & 0.17 & 0.13 & 0.12 & 0.15 & 0.17 & 0.19 & 0.18 & 0.18 \\
Velocity & - & - & - & - & - & - & - & 0.08 & 0.08 & 0.11 & 0.14 & 0.20 & 0.25 & 0.23 & 0.15 & \textbf{0.10} & 0.11 & 0.12 & 0.13 & 0.13 & 0.14 \\
Joint multi-task & 0.05 & 0.05 & 0.04 & 0.04 & 0.07 & \textbf{0.02} & 0.04 & 0.13 & 0.12 & 0.12 & 0.13 & 0.16 & 0.18 & 0.18 & 0.12 & \textbf{0.10} & 0.13 & 0.12 & 0.14 & 0.18 & 0.18 \\
Joint & 0.04 & 0.03 & 0.03 & 0.04 & 0.10 & 0.13 & 0.15 & 0.08 & 0.09 & 0.14 & 0.17 & 0.19 & 0.22 & 0.24 & \textbf{0.11} & 0.11 & 0.11 & \textbf{0.08} & \textbf{0.06} & \textbf{0.06} & \textbf{0.06} \\
\end{tabular}}

%% file: stats_rmse_cut_test.tex
\resizebox{1.7\textwidth}{!}{%
\begin{tabular}{c|ccccccc|ccccccc|ccccccc}
 & \multicolumn{7}{|c}{Chile} & \multicolumn{7}{|c}{Italy} & \multicolumn{7}{|c}{Japan} \\
 & 0.5s & 1.0s & 2.0s & 4.0s & 8.0s & 16.0s & 25.0s & 0.5s & 1.0s & 2.0s & 4.0s & 8.0s & 16.0s & 25.0s & 0.5s & 1.0s & 2.0s & 4.0s & 8.0s & 16.0s & 25.0s \\
\hline
Baseline & - & 0.95 & \textbf{0.71} & \textbf{0.58} & \textbf{0.49} & 0.45 & 0.52 & - & 1.32 & 1.18 & 0.86 & 0.53 & 0.51 & 0.50 & - & 1.09 & 0.73 & 0.57 & 0.54 & 0.58 & 0.50 \\
Multi-task & 1.38 & 1.06 & 0.83 & 0.67 & 0.60 & 0.46 & 0.41 & 1.12 & 1.04 & 0.90 & 0.81 & 0.68 & 0.67 & 0.66 & 1.31 & 1.09 & 0.90 & 0.91 & 0.69 & 0.40 & 0.38 \\
High magnitudes & \textbf{0.89} & \textbf{0.89} & 0.90 & 0.90 & 0.90 & 0.90 & 0.91 & \textbf{0.81} & \textbf{0.83} & 0.85 & 0.87 & 0.84 & 0.84 & 0.85 & \textbf{0.59} & \textbf{0.58} & \textbf{0.54} & \textbf{0.52} & \textbf{0.42} & \textbf{0.35} & 0.41 \\
Plain & 1.43 & 1.11 & 0.88 & 0.70 & 0.60 & 0.46 & 0.40 & 1.02 & 0.96 & 0.82 & 0.62 & 0.47 & 0.42 & 0.40 & 1.21 & 1.01 & 0.85 & 0.86 & 0.66 & 0.38 & 0.39 \\
 & $\pm$0.02 & $\pm$0.03 & $\pm$0.06 & $\pm$0.04 & $\pm$0.05 & $\pm$0.03 & $\pm$0.01 & $\pm$0.04 & $\pm$0.02 & $\pm$0.04 & $\pm$0.05 & $\pm$0.04 & $\pm$0.04 & $\pm$0.05 & $\pm$0.03 & $\pm$0.05 & $\pm$0.04 & $\pm$0.04 & $\pm$0.05 & $\pm$0.04 & $\pm$0.04 \\
Ensemble & 1.41 & 1.09 & 0.84 & 0.68 & 0.59 & 0.45 & 0.39 & 0.98 & 0.93 & 0.80 & 0.59 & 0.44 & 0.38 & 0.36 & 1.17 & 0.98 & 0.85 & 0.84 & 0.66 & \textbf{0.35} & \textbf{0.35} \\
No Upsampling & 1.56 & 1.30 & 1.05 & 0.82 & 0.74 & 0.56 & 0.51 & 1.03 & 0.91 & 0.80 & 0.64 & 0.48 & 0.43 & 0.43 & 1.46 & 1.22 & 1.02 & 0.98 & 0.78 & 0.52 & 0.50 \\
Pooling (Emb) & 1.40 & 1.11 & 0.88 & 0.68 & 0.55 & 0.42 & 0.37 & 1.02 & 0.94 & 0.82 & 0.62 & 0.49 & 0.45 & 0.44 & 1.23 & 1.00 & 0.87 & 0.87 & 0.71 & 0.44 & 0.41 \\
 & $\pm$0.02 & $\pm$0.07 & $\pm$0.03 & $\pm$0.03 & $\pm$0.03 & $\pm$0.03 & $\pm$0.02 & $\pm$0.12 & $\pm$0.12 & $\pm$0.11 & $\pm$0.13 & $\pm$0.13 & $\pm$0.15 & $\pm$0.15 & $\pm$0.04 & $\pm$0.04 & $\pm$0.05 & $\pm$0.08 & $\pm$0.05 & $\pm$0.06 & $\pm$0.06 \\
Pooling (Concat) & 1.37 & 1.05 & 0.88 & 0.68 & 0.54 & 0.42 & 0.39 & 0.96 & 0.88 & \textbf{0.75} & \textbf{0.56} & 0.46 & 0.40 & 0.40 & 1.19 & 0.98 & 0.84 & 0.83 & 0.66 & 0.39 & 0.39 \\
 & $\pm$0.02 & $\pm$0.06 & $\pm$0.02 & $\pm$0.06 & $\pm$0.06 & $\pm$0.04 & $\pm$0.04 & $\pm$0.04 & $\pm$0.04 & $\pm$0.03 & $\pm$0.04 & $\pm$0.04 & $\pm$0.03 & $\pm$0.04 & $\pm$0.03 & $\pm$0.03 & $\pm$0.03 & $\pm$0.03 & $\pm$0.03 & $\pm$0.01 & $\pm$0.01 \\
Transfer & 1.33 & 1.01 & 0.83 & 0.67 & 0.55 & \textbf{0.37} & \textbf{0.34} & 1.12 & 1.05 & 0.85 & 0.62 & 0.38 & 0.40 & 0.37 & 1.32 & 1.15 & 0.96 & 0.93 & 0.80 & 0.44 & 0.41 \\
Multi-task transfer & 1.32 & 1.02 & \textbf{0.74} & \textbf{0.61} & \textbf{0.50} & 0.42 & 0.37 & 1.13 & 1.02 & 0.82 & 0.61 & 0.48 & 0.46 & 0.48 & 1.45 & 1.10 & 0.93 & 0.87 & 0.63 & \textbf{0.36} & 0.38 \\
Velocity & - & - & - & - & - & - & - & 1.13 & 1.02 & 0.82 & 0.67 & 0.50 & 0.50 & 0.50 & 1.46 & 1.18 & 0.98 & 0.91 & 0.72 & 0.41 & 0.42 \\
Joint multi-task & 2.57 & 2.00 & 1.37 & 0.99 & 0.87 & 0.82 & 0.71 & 2.52 & 2.64 & 1.66 & 1.32 & 1.06 & 1.02 & 1.02 & 1.49 & 1.15 & 1.01 & 0.92 & 0.72 & \textbf{0.35} & \textbf{0.37} \\
Joint & 2.50 & 2.08 & 1.68 & 1.37 & 1.26 & 1.01 & 0.77 & 2.06 & 2.61 & 1.48 & 1.14 & \textbf{0.26} & \textbf{0.13} & \textbf{0.16} & 1.38 & 1.12 & 0.91 & 0.87 & 0.75 & 0.44 & 0.46 \\
\end{tabular}}

%% file: stats_mae_cut_test.tex
\resizebox{1.7\textwidth}{!}{%
\begin{tabular}{c|ccccccc|ccccccc|ccccccc}
 & \multicolumn{7}{|c}{Chile} & \multicolumn{7}{|c}{Italy} & \multicolumn{7}{|c}{Japan} \\
 & 0.5s & 1.0s & 2.0s & 4.0s & 8.0s & 16.0s & 25.0s & 0.5s & 1.0s & 2.0s & 4.0s & 8.0s & 16.0s & 25.0s & 0.5s & 1.0s & 2.0s & 4.0s & 8.0s & 16.0s & 25.0s \\
\hline
Baseline & - & 0.74 & 0.57 & 0.46 & \textbf{0.35} & 0.30 & 0.34 & - & 1.16 & 0.98 & 0.73 & 0.45 & 0.40 & 0.37 & - & 0.79 & 0.55 & 0.43 & 0.41 & 0.42 & 0.39 \\
Multi-task & 0.98 & 0.73 & 0.55 & 0.44 & 0.38 & 0.30 & 0.27 & 0.94 & 0.86 & 0.74 & 0.63 & 0.54 & 0.54 & 0.53 & 1.02 & 0.76 & 0.58 & 0.52 & 0.38 & 0.28 & 0.27 \\
High magnitudes & \textbf{0.58} & \textbf{0.58} & 0.58 & 0.57 & 0.58 & 0.58 & 0.58 & \textbf{0.63} & \textbf{0.67} & 0.68 & 0.70 & 0.69 & 0.69 & 0.70 & \textbf{0.44} & \textbf{0.43} & \textbf{0.40} & \textbf{0.38} & \textbf{0.30} & 0.27 & 0.30 \\
Plain & 1.05 & 0.78 & 0.60 & 0.45 & 0.38 & 0.29 & 0.25 & 0.81 & 0.74 & 0.61 & 0.43 & 0.34 & 0.30 & 0.29 & 0.94 & 0.73 & 0.57 & 0.48 & 0.38 & 0.26 & 0.27 \\
 & $\pm$0.04 & $\pm$0.03 & $\pm$0.04 & $\pm$0.04 & $\pm$0.02 & $\pm$0.02 & $\pm$0.02 & $\pm$0.05 & $\pm$0.04 & $\pm$0.04 & $\pm$0.06 & $\pm$0.05 & $\pm$0.05 & $\pm$0.06 & $\pm$0.03 & $\pm$0.04 & $\pm$0.03 & $\pm$0.04 & $\pm$0.03 & $\pm$0.03 & $\pm$0.03 \\
Ensemble & 1.04 & 0.76 & 0.57 & 0.44 & 0.38 & 0.29 & 0.24 & 0.75 & 0.71 & 0.59 & 0.40 & 0.30 & 0.26 & 0.25 & 0.90 & 0.69 & 0.53 & 0.45 & 0.36 & \textbf{0.24} & \textbf{0.24} \\
No Upsampling & 1.19 & 0.97 & 0.73 & 0.55 & 0.44 & 0.37 & 0.33 & 0.81 & \textbf{0.68} & \textbf{0.57} & 0.42 & 0.33 & 0.29 & 0.29 & 1.17 & 0.95 & 0.75 & 0.63 & 0.47 & 0.37 & 0.36 \\
Pooling (Emb) & 1.00 & 0.77 & 0.59 & 0.44 & \textbf{0.35} & 0.27 & 0.23 & 0.81 & 0.73 & 0.63 & 0.44 & 0.36 & 0.33 & 0.33 & 0.95 & 0.74 & 0.59 & 0.53 & 0.43 & 0.33 & 0.31 \\
 & $\pm$0.02 & $\pm$0.05 & $\pm$0.02 & $\pm$0.02 & $\pm$0.02 & $\pm$0.02 & $\pm$0.02 & $\pm$0.16 & $\pm$0.16 & $\pm$0.14 & $\pm$0.15 & $\pm$0.15 & $\pm$0.15 & $\pm$0.16 & $\pm$0.03 & $\pm$0.04 & $\pm$0.06 & $\pm$0.08 & $\pm$0.06 & $\pm$0.07 & $\pm$0.06 \\
Pooling (Concat) & 0.98 & 0.74 & 0.59 & 0.45 & \textbf{0.36} & 0.28 & 0.26 & 0.74 & \textbf{0.66} & \textbf{0.55} & \textbf{0.38} & 0.31 & 0.28 & 0.28 & 0.92 & 0.72 & 0.57 & 0.50 & 0.38 & 0.27 & 0.26 \\
 & $\pm$0.02 & $\pm$0.05 & $\pm$0.03 & $\pm$0.06 & $\pm$0.04 & $\pm$0.04 & $\pm$0.03 & $\pm$0.03 & $\pm$0.03 & $\pm$0.03 & $\pm$0.02 & $\pm$0.02 & $\pm$0.02 & $\pm$0.03 & $\pm$0.03 & $\pm$0.04 & $\pm$0.03 & $\pm$0.02 & $\pm$0.02 & $\pm$0.01 & $\pm$0.01 \\
Transfer & 0.98 & 0.70 & 0.55 & 0.45 & \textbf{0.34} & \textbf{0.23} & \textbf{0.21} & 0.93 & 0.87 & 0.62 & \textbf{0.40} & \textbf{0.23} & 0.25 & 0.24 & 0.94 & 0.76 & 0.62 & 0.50 & 0.41 & 0.29 & 0.27 \\
Multi-task transfer & 0.94 & 0.70 & \textbf{0.48} & \textbf{0.40} & \textbf{0.35} & 0.29 & 0.24 & 0.91 & 0.80 & 0.62 & 0.44 & 0.36 & 0.34 & 0.35 & 1.05 & 0.72 & 0.55 & 0.49 & 0.34 & 0.25 & 0.27 \\
Velocity & - & - & - & - & - & - & - & 0.95 & 0.82 & 0.63 & 0.51 & 0.39 & 0.38 & 0.38 & 1.11 & 0.87 & 0.63 & 0.52 & 0.41 & 0.26 & 0.26 \\
Joint multi-task & 2.42 & 1.76 & 1.09 & 0.76 & 0.69 & 0.65 & 0.53 & 2.46 & 2.60 & 1.64 & 1.30 & 0.99 & 0.95 & 0.95 & 1.09 & 0.76 & 0.63 & 0.51 & 0.37 & \textbf{0.24} & 0.26 \\
Joint & 2.40 & 1.94 & 1.36 & 1.11 & 1.04 & 0.81 & 0.59 & 1.80 & 2.61 & 1.36 & 1.14 & 0.26 & \textbf{0.11} & \textbf{0.16} & 1.04 & 0.77 & 0.60 & 0.50 & 0.39 & 0.31 & 0.32 \\
\end{tabular}}

%% file: stats_r2_score_cut_test.tex
\resizebox{1.7\textwidth}{!}{%
\begin{tabular}{c|ccccccc|ccccccc|ccccccc}
 & \multicolumn{7}{|c}{Chile} & \multicolumn{7}{|c}{Italy} & \multicolumn{7}{|c}{Japan} \\
 & 0.5s & 1.0s & 2.0s & 4.0s & 8.0s & 16.0s & 25.0s & 0.5s & 1.0s & 2.0s & 4.0s & 8.0s & 16.0s & 25.0s & 0.5s & 1.0s & 2.0s & 4.0s & 8.0s & 16.0s & 25.0s \\
\hline
Baseline & - & -0.95 & -0.09 & \textbf{0.26} & \textbf{0.49} & 0.57 & 0.41 & - & -4.15 & -4.46 & -1.77 & -0.09 & -0.21 & -0.67 & - & -4.21 & -1.37 & -0.47 & -0.41 & -0.55 & -0.16 \\
Multi-task & -3.14 & -1.45 & -0.51 & 0.04 & 0.22 & 0.55 & 0.63 & -3.99 & -3.27 & -2.24 & -1.58 & -0.83 & -0.78 & -0.72 & -7.01 & -4.52 & -2.80 & -2.83 & -1.22 & 0.27 & 0.31 \\
High magnitudes & -0.45 & -0.45 & -0.48 & -0.46 & -0.47 & -0.47 & -0.50 & -1.62 & -1.76 & -1.85 & -2.03 & -1.83 & -1.80 & -1.84 & -0.41 & -0.36 & -0.18 & -0.12 & \textbf{0.27} & \textbf{0.48} & 0.31 \\
Plain & -3.41 & -1.69 & -0.71 & -0.05 & 0.20 & 0.53 & 0.65 & -3.14 & -2.67 & -1.67 & -0.52 & 0.13 & 0.29 & 0.34 & -5.83 & -3.77 & -2.37 & -2.42 & -1.07 & 0.30 & 0.27 \\
 & $\pm$0.15 & $\pm$0.13 & $\pm$0.24 & $\pm$0.12 & $\pm$0.12 & $\pm$0.06 & $\pm$0.02 & $\pm$0.29 & $\pm$0.17 & $\pm$0.25 & $\pm$0.23 & $\pm$0.17 & $\pm$0.15 & $\pm$0.16 & $\pm$0.35 & $\pm$0.49 & $\pm$0.32 & $\pm$0.30 & $\pm$0.33 & $\pm$0.13 & $\pm$0.13 \\
Ensemble & -3.31 & -1.57 & -0.55 & 0.00 & 0.24 & 0.55 & 0.67 & -2.85 & -2.44 & -1.53 & -0.41 & 0.25 & 0.43 & 0.49 & -5.36 & -3.51 & -2.32 & -2.30 & -1.01 & 0.44 & \textbf{0.42} \\
No Upsampling & -4.26 & -2.68 & -1.39 & -0.44 & -0.19 & 0.32 & 0.44 & -3.18 & -2.30 & -1.55 & -0.62 & 0.08 & 0.28 & 0.27 & -8.93 & -5.95 & -3.88 & -3.48 & -1.81 & -0.25 & -0.17 \\
Pooling (Emb) & -3.24 & -1.69 & -0.69 & -0.02 & 0.35 & 0.61 & 0.71 & -3.18 & -2.59 & -1.69 & -0.60 & -0.03 & 0.13 & 0.15 & -6.01 & -3.69 & -2.55 & -2.52 & -1.36 & 0.10 & 0.19 \\
 & $\pm$0.12 & $\pm$0.36 & $\pm$0.12 & $\pm$0.10 & $\pm$0.07 & $\pm$0.06 & $\pm$0.04 & $\pm$1.07 & $\pm$0.98 & $\pm$0.73 & $\pm$0.70 & $\pm$0.63 & $\pm$0.64 & $\pm$0.66 & $\pm$0.51 & $\pm$0.42 & $\pm$0.45 & $\pm$0.65 & $\pm$0.32 & $\pm$0.27 & $\pm$0.22 \\
Pooling (Concat) & -3.06 & -1.41 & -0.69 & -0.02 & 0.35 & 0.62 & 0.66 & -2.64 & -2.09 & -1.21 & -0.26 & 0.17 & 0.35 & 0.37 & -5.53 & -3.47 & -2.31 & -2.24 & -1.03 & 0.28 & 0.29 \\
 & $\pm$0.12 & $\pm$0.27 & $\pm$0.08 & $\pm$0.17 & $\pm$0.15 & $\pm$0.08 & $\pm$0.06 & $\pm$0.31 & $\pm$0.25 & $\pm$0.20 & $\pm$0.17 & $\pm$0.14 & $\pm$0.11 & $\pm$0.12 & $\pm$0.37 & $\pm$0.26 & $\pm$0.27 & $\pm$0.23 & $\pm$0.21 & $\pm$0.04 & $\pm$0.05 \\
Transfer & -2.82 & -1.22 & -0.48 & 0.03 & 0.36 & \textbf{0.70} & \textbf{0.75} & -3.95 & -3.39 & -1.89 & -0.51 & \textbf{0.43} & 0.37 & 0.44 & -7.15 & -5.19 & -3.32 & -3.03 & -1.99 & 0.11 & 0.21 \\
Multi-task transfer & -2.79 & -1.27 & -0.19 & 0.21 & 0.45 & 0.61 & 0.70 & -4.11 & -3.14 & -1.69 & -0.50 & 0.08 & 0.15 & 0.10 & -8.72 & -4.65 & -3.03 & -2.53 & -0.83 & 0.39 & 0.34 \\
Velocity & - & - & - & - & - & - & - & -4.04 & -3.17 & -1.67 & -0.79 & 0.01 & 0.01 & 0.01 & -8.92 & -5.50 & -3.44 & -2.87 & -1.39 & 0.22 & 0.18 \\
Joint multi-task & -13.69 & -7.88 & -3.17 & -1.18 & -0.67 & -0.51 & -0.10 & -100.77 & -110.26 & -42.87 & -26.91 & -16.89 & -15.79 & -15.52 & -9.29 & -5.10 & -3.71 & -2.97 & -1.44 & 0.43 & 0.38 \\
Joint & -12.85 & -8.63 & -5.23 & -3.17 & -2.53 & -1.28 & -0.32 & -66.63 & -108.03 & -34.05 & -19.94 & -0.08 & \textbf{0.74} & \textbf{0.59} & -7.90 & -4.79 & -2.89 & -2.53 & -1.58 & 0.10 & 0.04 \\
\end{tabular}}

%% file: loc_stats_rmse_hypo_test.tex
\resizebox{1.7\textwidth}{!}{%
\begin{tabular}{c|ccccccc|ccccccc|ccccccc}
 & \multicolumn{7}{|c}{Chile} & \multicolumn{7}{|c}{Italy} & \multicolumn{7}{|c}{Japan} \\
 & 0.5s & 1.0s & 2.0s & 4.0s & 8.0s & 16.0s & 25.0s & 0.5s & 1.0s & 2.0s & 4.0s & 8.0s & 16.0s & 25.0s & 0.5s & 1.0s & 2.0s & 4.0s & 8.0s & 16.0s & 25.0s \\
\hline
Baseline & 78.97 & 77.62 & 74.32 & 69.53 & 44.28 & 17.51 & 17.58 & 66.92 & 75.15 & 91.77 & 189.05 & 172.65 & 69.88 & 56.09 & 82.89 & 88.03 & 90.80 & 98.55 & 184.61 & 373.34 & 332.11 \\
Multi-task & 48.49 & 40.17 & 33.10 & 24.79 & 17.69 & 12.63 & 11.86 & 67.06 & 65.52 & 61.91 & 58.09 & 55.13 & 54.93 & 56.25 & 140.83 & 141.21 & 144.12 & 139.64 & 135.31 & 134.77 & 134.13 \\
High magnitudes & 124.94 & 113.56 & 113.02 & 119.20 & 116.61 & 90.28 & 93.75 & 192.27 & 190.97 & 189.82 & 188.28 & 186.77 & 187.36 & 188.10 & 393.78 & 377.11 & 365.31 & 383.16 & 352.42 & 344.42 & 352.45 \\
Plain & 43.96 & 36.99 & 28.66 & 20.74 & 12.52 & 8.97 & 8.83 & \textbf{51.94} & \textbf{50.35} & \textbf{45.49} & \textbf{41.07} & \textbf{35.33} & \textbf{35.89} & \textbf{38.68} & 77.08 & 73.51 & 72.16 & 69.93 & 68.18 & 66.60 & 66.10 \\
Ensemble & \textbf{39.36} & \textbf{32.31} & \textbf{25.04} & \textbf{18.24} & \textbf{10.77} & \textbf{7.90} & \textbf{7.81} & 63.28 & 61.57 & 57.77 & 54.52 & 51.95 & 52.00 & 53.38 & \textbf{64.15} & \textbf{61.48} & \textbf{60.09} & \textbf{58.64} & 56.45 & 54.26 & \textbf{54.07} \\
Pooling (Emb) & 44.04 & 37.00 & 30.42 & 25.30 & 18.56 & 15.10 & 14.40 & 65.73 & 65.00 & 61.25 & 59.02 & 56.91 & 57.07 & 58.21 & 98.95 & 95.17 & 93.88 & 91.29 & 88.20 & 87.26 & 87.13 \\
Pooling (Concat) & 53.19 & 44.53 & 38.02 & 33.01 & 27.09 & 22.31 & 21.80 & 181.39 & 181.41 & 181.47 & 181.52 & 181.46 & 181.47 & 181.52 & 163.43 & 155.61 & 154.13 & 152.15 & 149.51 & 148.23 & 148.29 \\
Transfer & 44.92 & 37.96 & 30.76 & 21.24 & 13.78 & 9.05 & 9.25 & \textbf{52.63} & \textbf{51.21} & \textbf{44.56} & \textbf{41.04} & \textbf{36.99} & \textbf{36.73} & \textbf{39.38} & \textbf{64.80} & \textbf{60.76} & \textbf{60.72} & \textbf{56.79} & \textbf{53.61} & \textbf{51.32} & \textbf{52.27} \\
Multi-task transfer & 51.07 & 41.46 & 32.91 & 23.00 & 14.56 & 10.80 & 10.46 & 55.78 & 54.26 & 50.60 & 47.10 & 43.82 & 45.37 & 49.68 & 98.96 & 91.05 & 90.85 & 87.61 & 86.29 & 87.75 & 88.27 \\
Velocity & - & - & - & - & - & - & - & 75.50 & 74.75 & 71.02 & 68.44 & 66.97 & 66.67 & 68.56 & 77.41 & 69.06 & 67.91 & 64.79 & 62.96 & 61.95 & 60.64 \\
Joint multi-task & 52.36 & 43.38 & 35.14 & 25.23 & 17.32 & 14.82 & 14.64 & 62.87 & 61.27 & 56.59 & 54.38 & 51.56 & 51.81 & 53.81 & 111.41 & 100.82 & 100.55 & 97.64 & 97.20 & 97.92 & 97.39 \\
Joint & 49.55 & 42.75 & 35.52 & 24.32 & 15.69 & 10.12 & 9.76 & 67.38 & 65.06 & 60.95 & 58.38 & 58.10 & 59.13 & 60.29 & 198.47 & 196.29 & 195.19 & 196.00 & 194.68 & 194.46 & 195.26 \\
\end{tabular}}

%% file: loc_stats_mae_hypo_test.tex
\resizebox{1.7\textwidth}{!}{%
\begin{tabular}{c|ccccccc|ccccccc|ccccccc}
 & \multicolumn{7}{|c}{Chile} & \multicolumn{7}{|c}{Italy} & \multicolumn{7}{|c}{Japan} \\
 & 0.5s & 1.0s & 2.0s & 4.0s & 8.0s & 16.0s & 25.0s & 0.5s & 1.0s & 2.0s & 4.0s & 8.0s & 16.0s & 25.0s & 0.5s & 1.0s & 2.0s & 4.0s & 8.0s & 16.0s & 25.0s \\
\hline
Baseline & 73.30 & 71.60 & 67.38 & 58.00 & 28.15 & 13.80 & 13.92 & 26.03 & 32.11 & 50.13 & 111.88 & 71.80 & 18.57 & 15.70 & 43.66 & 46.38 & 49.29 & 54.04 & 78.95 & 154.71 & 124.94 \\
Multi-task & 33.78 & 27.71 & 22.39 & 15.81 & 10.99 & 8.20 & 7.89 & 27.09 & 25.25 & 22.88 & 20.27 & 18.87 & 18.76 & 19.22 & 67.48 & 66.47 & 65.97 & 63.09 & 59.05 & 57.13 & 56.76 \\
High magnitudes & 106.43 & 98.18 & 99.49 & 103.49 & 104.08 & 82.75 & 85.67 & 120.07 & 119.24 & 119.07 & 117.63 & 116.63 & 116.50 & 116.93 & 229.31 & 221.00 & 218.40 & 218.67 & 208.64 & 203.95 & 207.15 \\
Plain & 30.41 & 25.01 & 19.06 & 12.40 & 7.26 & 5.42 & 5.30 & \textbf{20.47} & \textbf{18.84} & \textbf{16.59} & 14.22 & \textbf{11.86} & 11.85 & 12.91 & 40.88 & 37.35 & 35.43 & 33.15 & 30.28 & 29.20 & 29.06 \\
Ensemble & \textbf{27.20} & \textbf{22.09} & \textbf{16.94} & \textbf{11.06} & \textbf{6.55} & \textbf{4.90} & \textbf{4.77} & 28.88 & 27.68 & 25.78 & 24.20 & 23.11 & 23.09 & 23.48 & 35.25 & 31.59 & 29.49 & 27.50 & 24.47 & 23.11 & 22.88 \\
Pooling (Emb) & 30.75 & 25.44 & 20.86 & 16.72 & 12.61 & 9.60 & 9.04 & 28.16 & 26.93 & 24.85 & 23.37 & 22.20 & 22.23 & 22.59 & 51.57 & 45.70 & 43.15 & 40.34 & 37.48 & 35.94 & 35.75 \\
Pooling (Concat) & 37.68 & 30.62 & 25.96 & 21.94 & 17.89 & 14.11 & 13.68 & 83.53 & 83.47 & 83.47 & 83.58 & 83.51 & 83.42 & 83.43 & 125.56 & 118.22 & 117.21 & 116.03 & 113.29 & 111.72 & 111.59 \\
Transfer & 31.39 & 26.03 & 20.29 & 12.75 & 7.47 & 5.27 & 5.16 & \textbf{21.09} & \textbf{19.39} & \textbf{16.22} & \textbf{13.42} & \textbf{11.55} & \textbf{11.02} & \textbf{11.79} & \textbf{31.86} & \textbf{27.84} & \textbf{25.81} & \textbf{22.87} & \textbf{20.11} & \textbf{18.49} & \textbf{18.16} \\
Multi-task transfer & 35.86 & 28.93 & 22.46 & 14.69 & 9.10 & 6.77 & 6.60 & 24.66 & 23.51 & 21.80 & 19.75 & 18.74 & 19.00 & 20.18 & 48.75 & 43.28 & 40.07 & 37.48 & 35.21 & 34.39 & 34.21 \\
Velocity & - & - & - & - & - & - & - & 32.69 & 31.80 & 29.85 & 28.20 & 27.17 & 27.25 & 27.85 & 43.11 & 39.53 & 37.60 & 34.71 & 32.58 & 31.59 & 31.09 \\
Joint multi-task & 38.31 & 31.64 & 25.02 & 17.28 & 11.90 & 9.78 & 9.65 & 30.48 & 28.98 & 27.13 & 25.74 & 24.53 & 24.56 & 25.30 & 54.41 & 48.50 & 44.99 & 42.34 & 40.29 & 39.34 & 38.94 \\
Joint & 34.75 & 29.17 & 23.04 & 14.61 & 8.66 & 5.95 & 5.70 & 24.61 & 22.52 & 19.82 & 17.05 & 15.81 & 15.72 & 16.22 & 124.42 & 121.32 & 118.92 & 117.11 & 113.93 & 112.59 & 112.77 \\
\end{tabular}}

%% file: loc_stats_rmse_epi_test.tex
\resizebox{1.7\textwidth}{!}{%
\begin{tabular}{c|ccccccc|ccccccc|ccccccc}
 & \multicolumn{7}{|c}{Chile} & \multicolumn{7}{|c}{Italy} & \multicolumn{7}{|c}{Japan} \\
 & 0.5s & 1.0s & 2.0s & 4.0s & 8.0s & 16.0s & 25.0s & 0.5s & 1.0s & 2.0s & 4.0s & 8.0s & 16.0s & 25.0s & 0.5s & 1.0s & 2.0s & 4.0s & 8.0s & 16.0s & 25.0s \\
\hline
Baseline & 65.49 & 63.86 & 59.74 & 52.07 & 29.72 & 13.82 & 13.82 & 55.77 & 64.82 & 67.69 & 84.57 & 116.35 & 66.82 & 53.17 & 73.93 & 79.82 & 82.69 & 86.91 & 160.74 & 360.57 & 325.00 \\
Multi-task & 44.19 & 36.51 & 29.74 & 22.46 & 16.20 & 11.29 & 10.46 & 61.81 & 60.39 & 58.22 & 56.28 & 53.67 & 53.46 & 54.20 & 138.89 & 139.48 & 142.85 & 138.34 & 134.03 & 133.63 & 133.03 \\
High magnitudes & 121.57 & 111.06 & 110.52 & 115.92 & 114.20 & 88.24 & 90.81 & 187.58 & 186.33 & 185.25 & 183.85 & 182.34 & 182.95 & 183.69 & 387.59 & 367.30 & 352.89 & 375.67 & 338.82 & 331.22 & 338.71 \\
Plain & 39.48 & 32.95 & 25.25 & 18.34 & 11.24 & 7.73 & 7.58 & \textbf{45.62} & \textbf{44.65} & \textbf{40.95} & \textbf{38.06} & \textbf{33.15} & \textbf{33.51} & \textbf{35.20} & 73.10 & 70.09 & 69.37 & 67.40 & 66.22 & 64.86 & 64.30 \\
Ensemble & \textbf{35.68} & \textbf{29.21} & \textbf{22.28} & \textbf{16.12} & \textbf{9.30} & \textbf{6.70} & \textbf{6.64} & 58.05 & 57.09 & 53.99 & 52.09 & 49.95 & 49.92 & 50.59 & \textbf{59.82} & \textbf{57.60} & \textbf{56.75} & 55.52 & 53.94 & 52.15 & 51.98 \\
Pooling (Emb) & 40.11 & 33.68 & 27.41 & 22.72 & 16.53 & 13.79 & 13.27 & 59.18 & 58.42 & 54.67 & 52.83 & 50.91 & 51.20 & 52.29 & 95.42 & 91.99 & 91.39 & 88.92 & 86.01 & 85.25 & 85.14 \\
Pooling (Concat) & 49.57 & 41.56 & 35.44 & 30.80 & 25.52 & 21.06 & 20.77 & 178.40 & 178.41 & 178.43 & 178.46 & 178.43 & 178.45 & 178.51 & 161.48 & 153.83 & 152.69 & 151.03 & 148.65 & 147.48 & 147.56 \\
Transfer & 39.80 & 33.04 & 26.20 & 18.72 & 12.51 & 7.95 & 8.21 & \textbf{46.87} & \textbf{45.63} & \textbf{41.84} & \textbf{39.46} & 35.66 & 35.80 & 37.94 & \textbf{59.28} & \textbf{55.24} & \textbf{55.34} & \textbf{52.53} & \textbf{50.00} & \textbf{48.05} & \textbf{49.31} \\
Multi-task transfer & 46.17 & 37.03 & 28.98 & 20.61 & 12.97 & 9.45 & 9.12 & 49.79 & 48.63 & 45.65 & 43.73 & 40.68 & 42.06 & 45.25 & 95.24 & 87.75 & 87.95 & 84.75 & 83.85 & 85.67 & 86.23 \\
Velocity & - & - & - & - & - & - & - & 70.61 & 70.14 & 66.79 & 65.30 & 63.80 & 63.76 & 64.77 & 72.34 & 64.27 & 63.83 & 60.69 & 59.33 & 58.74 & 57.48 \\
Joint multi-task & 47.10 & 38.58 & 30.60 & 22.14 & 14.93 & 13.05 & 12.80 & 56.11 & 54.73 & 50.60 & 48.94 & 46.58 & 47.09 & 48.61 & 108.26 & 97.65 & 97.53 & 94.64 & 94.43 & 95.36 & 94.80 \\
Joint & 44.46 & 37.75 & 30.83 & 21.41 & 14.44 & 8.86 & 8.53 & 62.76 & 61.32 & 58.91 & 57.21 & 57.06 & 57.98 & 58.71 & 197.03 & 195.09 & 194.19 & 195.16 & 194.27 & 194.13 & 194.85 \\
\end{tabular}}

%% file: loc_stats_mae_epi_test.tex
\resizebox{1.7\textwidth}{!}{%
\begin{tabular}{c|ccccccc|ccccccc|ccccccc}
 & \multicolumn{7}{|c}{Chile} & \multicolumn{7}{|c}{Italy} & \multicolumn{7}{|c}{Japan} \\
 & 0.5s & 1.0s & 2.0s & 4.0s & 8.0s & 16.0s & 25.0s & 0.5s & 1.0s & 2.0s & 4.0s & 8.0s & 16.0s & 25.0s & 0.5s & 1.0s & 2.0s & 4.0s & 8.0s & 16.0s & 25.0s \\
\hline
Baseline & 58.27 & 56.23 & 51.33 & 40.82 & 18.73 & 10.25 & 10.28 & 21.58 & 26.72 & 33.11 & 45.59 & 46.45 & 14.30 & \textbf{10.58} & 34.11 & 37.77 & 41.37 & 45.18 & 64.04 & 141.80 & 114.48 \\
Multi-task & 30.01 & 24.41 & 19.21 & 12.94 & 8.79 & 6.38 & 6.05 & 24.34 & 22.67 & 20.70 & 18.78 & 17.57 & 17.47 & 17.52 & 64.06 & 63.41 & 63.32 & 60.74 & 57.21 & 55.60 & 55.31 \\
High magnitudes & 103.82 & 96.17 & 97.49 & 100.94 & 102.08 & 80.90 & 83.17 & 116.66 & 115.99 & 115.90 & 114.44 & 113.44 & 113.30 & 113.71 & 219.98 & 209.35 & 206.58 & 207.45 & 196.24 & 191.91 & 195.00 \\
Plain & 26.76 & 21.67 & 16.00 & 9.63 & 5.52 & 3.96 & 3.86 & \textbf{17.45} & \textbf{16.05} & \textbf{14.20} & \textbf{12.38} & \textbf{10.23} & \textbf{10.19} & \textbf{11.01} & 37.39 & 34.24 & 32.54 & 30.44 & 28.43 & 27.65 & 27.50 \\
Ensemble & \textbf{23.74} & \textbf{18.94} & \textbf{13.98} & \textbf{8.33} & \textbf{4.85} & \textbf{3.56} & \textbf{3.44} & 25.85 & 24.95 & 23.36 & 22.24 & 21.42 & 21.38 & 21.36 & 31.78 & 28.46 & 26.57 & 24.99 & 22.89 & 21.76 & 21.51 \\
Pooling (Emb) & 27.15 & 22.27 & 17.82 & 13.70 & 10.32 & 7.93 & 7.45 & 24.01 & 22.75 & 20.68 & 19.47 & 18.48 & 18.54 & 18.79 & 47.74 & 42.31 & 40.22 & 37.77 & 35.36 & 34.04 & 33.88 \\
Pooling (Concat) & 34.09 & 27.53 & 23.11 & 19.18 & 15.77 & 12.50 & 12.11 & 82.19 & 82.10 & 82.01 & 82.02 & 82.02 & 81.96 & 81.98 & 122.65 & 116.01 & 115.44 & 114.68 & 112.38 & 111.04 & 110.94 \\
Transfer & 27.44 & 22.33 & 16.81 & 9.94 & 5.77 & 3.93 & 3.84 & 18.38 & \textbf{16.85} & \textbf{14.44} & \textbf{12.30} & \textbf{10.55} & \textbf{10.12} & \textbf{10.67} & \textbf{27.90} & \textbf{23.94} & \textbf{22.08} & \textbf{19.80} & \textbf{18.12} & \textbf{16.67} & \textbf{16.36} \\
Multi-task transfer & 31.86 & 25.37 & 19.18 & 11.73 & 7.09 & 5.17 & 5.02 & 20.92 & 19.86 & 18.63 & 17.33 & 16.53 & 16.71 & 17.45 & 44.96 & 39.92 & 37.04 & 34.72 & 33.03 & 32.51 & 32.31 \\
Velocity & - & - & - & - & - & - & - & 29.32 & 28.63 & 26.89 & 25.71 & 24.69 & 24.93 & 25.05 & 39.14 & 35.83 & 34.14 & 31.43 & 30.23 & 29.57 & 29.07 \\
Joint multi-task & 34.01 & 27.81 & 21.39 & 13.99 & 9.18 & 7.56 & 7.42 & 25.12 & 23.68 & 22.14 & 21.25 & 20.34 & 20.43 & 20.83 & 50.75 & 45.09 & 41.85 & 39.56 & 37.90 & 37.11 & 36.70 \\
Joint & 30.62 & 25.28 & 19.39 & 11.55 & 6.78 & 4.44 & 4.20 & 21.85 & 20.16 & 18.04 & 15.83 & 14.63 & 14.58 & 14.76 & 120.16 & 117.03 & 115.00 & 113.76 & 112.04 & 110.99 & 111.15 \\
\end{tabular}}